\documentclass[twocolumn,aps,tightenlinefs,superscriptaddress,preprintnumbers]{revtex4}
\usepackage{graphicx,color,dcolumn,booktabs,bm}
\usepackage{longtable,lscape}
\usepackage{amsmath}
\usepackage{indentfirst}
\usepackage{epsfig}
\usepackage{feynmf}   
\usepackage{epstopdf}   
\usepackage{slashed}  
\usepackage{cases}
\usepackage{color}
\usepackage{multirow}
\usepackage{ulem}
\usepackage{graphicx,color,dcolumn,booktabs,bm}
\usepackage{verbatim}
\usepackage[colorlinks,linkcolor=black,anchorcolor=green,citecolor=blue]{hyperref}
\usepackage{mathrsfs}
\usepackage{rotating}
\usepackage{threeparttable}
\usepackage{dcolumn}
\newcommand {\tabincell}[2]{\begin{tabular}{@{}#1@{}}#2\end{tabular}}%

\newcommand{\lum}{{\cal L}}

\newcommand{\BR}{{\cal B}}

\newcommand{\pip}{\pi^+}

\newcommand{\EE}{e^+e^-}

\newcommand{\KK}{K^+K^-}

\newcommand{\beq}{\begin{equation}}
\newcommand{\eeq}{\end{equation}}
\newcommand{\bitm}{\begin{itemize}}
\newcommand{\eitm}{\end{itemize}}

\newcommand{\ones}{\Upsilon(1S)}
\newcommand{\twos}{\Upsilon(2S)}
\newcommand{\onetwos}{\Upsilon(1S,2S)}

\newcommand{\fours}{\Upsilon(4S)}
\newcommand{\fives}{\Upsilon(5S)}

\newcommand{\dsds}{D_{s}^{+}D_{s}^{+}}
\newcommand{\dstdst}{D_{s}^{*+}D_{s}^{*+}}

\newcommand{\ccss}{X_{cc\bar{s}\bar{s}}}

\newcolumntype{d}[1]{D{.}{.}{#1}}

\begin{document}
\hyphenpenalty=5000
\tolerance=1000

\title{\quad\\[0.1cm]\boldmath Search for tetraquark states $X_{cc\bar{s}\bar{s}}$ in $D_{s}^{+}D_{s}^{+}~(D_{s}^{*+}D_{s}^{*+})$ final states at Belle}

\noaffiliation
\affiliation{Department of Physics, University of the Basque Country UPV/EHU, 48080 Bilbao}
\affiliation{University of Bonn, 53115 Bonn}
\affiliation{Brookhaven National Laboratory, Upton, New York 11973}
\affiliation{Budker Institute of Nuclear Physics SB RAS, Novosibirsk 630090}
\affiliation{Faculty of Mathematics and Physics, Charles University, 121 16 Prague}
\affiliation{Chonnam National University, Gwangju 61186}
\affiliation{Chung-Ang University, Seoul 06974}
\affiliation{University of Cincinnati, Cincinnati, Ohio 45221}
\affiliation{Deutsches Elektronen--Synchrotron, 22607 Hamburg}
\affiliation{Duke University, Durham, North Carolina 27708}
\affiliation{Institute of Theoretical and Applied Research (ITAR), Duy Tan University, Hanoi 100000}
\affiliation{Department of Physics, Fu Jen Catholic University, Taipei 24205}
\affiliation{Key Laboratory of Nuclear Physics and Ion-beam Application (MOE) and Institute of Modern Physics, Fudan University, Shanghai 200443}
\affiliation{Justus-Liebig-Universit\"at Gie\ss{}en, 35392 Gie\ss{}en}
\affiliation{Gifu University, Gifu 501-1193}
\affiliation{II. Physikalisches Institut, Georg-August-Universit\"at G\"ottingen, 37073 G\"ottingen}
\affiliation{SOKENDAI (The Graduate University for Advanced Studies), Hayama 240-0193}
\affiliation{Gyeongsang National University, Jinju 52828}
\affiliation{Department of Physics and Institute of Natural Sciences, Hanyang University, Seoul 04763}
\affiliation{University of Hawaii, Honolulu, Hawaii 96822}
\affiliation{High Energy Accelerator Research Organization (KEK), Tsukuba 305-0801}
\affiliation{J-PARC Branch, KEK Theory Center, High Energy Accelerator Research Organization (KEK), Tsukuba 305-0801}
\affiliation{National Research University Higher School of Economics, Moscow 101000}
\affiliation{Forschungszentrum J\"{u}lich, 52425 J\"{u}lich}
\affiliation{IKERBASQUE, Basque Foundation for Science, 48013 Bilbao}
\affiliation{Indian Institute of Science Education and Research Mohali, SAS Nagar, 140306}
\affiliation{Indian Institute of Technology Guwahati, Assam 781039}
\affiliation{Indian Institute of Technology Hyderabad, Telangana 502285}
\affiliation{Indian Institute of Technology Madras, Chennai 600036}
\affiliation{Indiana University, Bloomington, Indiana 47408}
\affiliation{Institute of High Energy Physics, Chinese Academy of Sciences, Beijing 100049}
\affiliation{Institute of High Energy Physics, Vienna 1050}
\affiliation{Institute for High Energy Physics, Protvino 142281}
\affiliation{INFN - Sezione di Napoli, I-80126 Napoli}
\affiliation{INFN - Sezione di Roma Tre, I-00146 Roma}
\affiliation{INFN - Sezione di Torino, I-10125 Torino}
\affiliation{Iowa State University, Ames, Iowa 50011}
\affiliation{Advanced Science Research Center, Japan Atomic Energy Agency, Naka 319-1195}
\affiliation{J. Stefan Institute, 1000 Ljubljana}
\affiliation{Institut f\"ur Experimentelle Teilchenphysik, Karlsruher Institut f\"ur Technologie, 76131 Karlsruhe}
\affiliation{Kavli Institute for the Physics and Mathematics of the Universe (WPI), University of Tokyo, Kashiwa 277-8583}
\affiliation{Kitasato University, Sagamihara 252-0373}
\affiliation{Korea Institute of Science and Technology Information, Daejeon 34141}
\affiliation{Korea University, Seoul 02841}
\affiliation{Kyoto Sangyo University, Kyoto 603-8555}
\affiliation{Kyungpook National University, Daegu 41566}
\affiliation{P.N. Lebedev Physical Institute of the Russian Academy of Sciences, Moscow 119991}
\affiliation{Liaoning Normal University, Dalian 116029}
\affiliation{Faculty of Mathematics and Physics, University of Ljubljana, 1000 Ljubljana}
\affiliation{Ludwig Maximilians University, 80539 Munich}
\affiliation{Luther College, Decorah, Iowa 52101}
\affiliation{Malaviya National Institute of Technology Jaipur, Jaipur 302017}
\affiliation{Faculty of Chemistry and Chemical Engineering, University of Maribor, 2000 Maribor}
\affiliation{Max-Planck-Institut f\"ur Physik, 80805 M\"unchen}
\affiliation{School of Physics, University of Melbourne, Victoria 3010}
\affiliation{University of Mississippi, University, Mississippi 38677}
\affiliation{University of Miyazaki, Miyazaki 889-2192}
\affiliation{Moscow Physical Engineering Institute, Moscow 115409}
\affiliation{Graduate School of Science, Nagoya University, Nagoya 464-8602}
\affiliation{Kobayashi-Maskawa Institute, Nagoya University, Nagoya 464-8602}
\affiliation{Universit\`{a} di Napoli Federico II, I-80126 Napoli}
\affiliation{Nara Women's University, Nara 630-8506}
\affiliation{National Central University, Chung-li 32054}
\affiliation{Department of Physics, National Taiwan University, Taipei 10617}
\affiliation{H. Niewodniczanski Institute of Nuclear Physics, Krakow 31-342}
\affiliation{Nippon Dental University, Niigata 951-8580}
\affiliation{Niigata University, Niigata 950-2181}
\affiliation{University of Nova Gorica, 5000 Nova Gorica}
\affiliation{Novosibirsk State University, Novosibirsk 630090}
\affiliation{Osaka City University, Osaka 558-8585}
\affiliation{Pacific Northwest National Laboratory, Richland, Washington 99352}
\affiliation{Panjab University, Chandigarh 160014}
\affiliation{University of Pittsburgh, Pittsburgh, Pennsylvania 15260}
\affiliation{Punjab Agricultural University, Ludhiana 141004}
\affiliation{Research Center for Nuclear Physics, Osaka University, Osaka 567-0047}
\affiliation{Meson Science Laboratory, Cluster for Pioneering Research, RIKEN, Saitama 351-0198}
\affiliation{Dipartimento di Matematica e Fisica, Universit\`{a} di Roma Tre, I-00146 Roma}
\affiliation{Department of Modern Physics and State Key Laboratory of Particle Detection and Electronics, University of Science and Technology of China, Hefei 230026}
\affiliation{Showa Pharmaceutical University, Tokyo 194-8543}
\affiliation{Soochow University, Suzhou 215006}
\affiliation{Soongsil University, Seoul 06978}
\affiliation{Sungkyunkwan University, Suwon 16419}
\affiliation{School of Physics, University of Sydney, New South Wales 2006}
\affiliation{Department of Physics, Faculty of Science, University of Tabuk, Tabuk 71451}
\affiliation{Tata Institute of Fundamental Research, Mumbai 400005}
\affiliation{Department of Physics, Technische Universit\"at M\"unchen, 85748 Garching}
\affiliation{Toho University, Funabashi 274-8510}
\affiliation{Department of Physics, Tohoku University, Sendai 980-8578}
\affiliation{Earthquake Research Institute, University of Tokyo, Tokyo 113-0032}
\affiliation{Department of Physics, University of Tokyo, Tokyo 113-0033}
\affiliation{Tokyo Institute of Technology, Tokyo 152-8550}
\affiliation{Tokyo Metropolitan University, Tokyo 192-0397}
\affiliation{Virginia Polytechnic Institute and State University, Blacksburg, Virginia 24061}
\affiliation{Wayne State University, Detroit, Michigan 48202}
\affiliation{Yamagata University, Yamagata 990-8560}
\affiliation{Yonsei University, Seoul 03722}
  \author{X.~Y.~Gao}\affiliation{Key Laboratory of Nuclear Physics and Ion-beam Application (MOE) and Institute of Modern Physics, Fudan University, Shanghai 200443} 
  \author{Y.~Li}\affiliation{Key Laboratory of Nuclear Physics and Ion-beam Application (MOE) and Institute of Modern Physics, Fudan University, Shanghai 200443} 
  \author{C.~P.~Shen}\affiliation{Key Laboratory of Nuclear Physics and Ion-beam Application (MOE) and Institute of Modern Physics, Fudan University, Shanghai 200443} 
  \author{I.~Adachi}\affiliation{High Energy Accelerator Research Organization (KEK), Tsukuba 305-0801}\affiliation{SOKENDAI (The Graduate University for Advanced Studies), Hayama 240-0193} 
  \author{H.~Aihara}\affiliation{Department of Physics, University of Tokyo, Tokyo 113-0033} 
  \author{D.~M.~Asner}\affiliation{Brookhaven National Laboratory, Upton, New York 11973} 
  \author{H.~Atmacan}\affiliation{University of Cincinnati, Cincinnati, Ohio 45221} 
  \author{T.~Aushev}\affiliation{National Research University Higher School of Economics, Moscow 101000} 
  \author{R.~Ayad}\affiliation{Department of Physics, Faculty of Science, University of Tabuk, Tabuk 71451} 
  \author{P.~Behera}\affiliation{Indian Institute of Technology Madras, Chennai 600036} 
% ...
\collaboration{The Belle Collaboration}


\begin{abstract}
A search for double-heavy tetraquark state candidates $\ccss$ decaying to $D_{s}^{+}D_{s}^{+}$ and $D_{s}^{*+} D_{s}^{*+}$ is presented for the first time using the data
samples of 102 million $\ones$ and 158 million $\twos$ events,
and the data samples at $\sqrt{s}$ = 10.52~GeV, 10.58~GeV, and 10.867~GeV corresponding to integrated luminosities of 89.5~fb$^{-1}$,
711.0~fb$^{-1}$, and 121.4~fb$^{-1}$, respectively, accumulated with the Belle detector at the KEKB asymmetric
energy electron-positron collider.
The invariant-mass spectra of the $D_{s}^{+}D_{s}^{+}$ and $D_{s}^{*+} D_{s}^{*+}$
are studied to search for possible resonances.
No significant signals are observed, and the 90\% confidence level upper limits on the product branching fractions
[$\BR(\Upsilon(1S,2S) \to X_{cc\bar{s}\bar{s}} + anything) \times \BR(X_{cc\bar{s}\bar{s}} \to D_{s}^{+}D_{s}^{+}(D_{s}^{*+} D_{s}^{*+}))$]
in $\onetwos$ inclusive decays and the product values of Born cross section and branching fraction
[$\sigma(\EE \to X_{cc\bar{s}\bar{s}} + anything ) \times \BR(X_{cc\bar{s}\bar{s}} \to D_{s}^{+}D_{s}^{+}(D_{s}^{*+} D_{s}^{*+}))$]
in $\EE$ collisions at $\sqrt{s}$ = 10.52~GeV, 10.58~GeV, and 10.867~GeV under different assumptions of $\ccss$ masses and widths are obtained.
\end{abstract}

\maketitle
\section{\boldmath Introduction}

The hadron spectrum was successfully categorized based on the quark model as early as the 1960s~\cite{quarkmodel}. For a long time, all known hadrons could be classified as mesons or baryons with components of a quark-antiquark pair ($q \bar{q}$) or three quarks ($qqq$), respectively. However, Quantum Chromodynamics (QCD) also allows the existence of more complex structures,
such as the tetraquark, pentaquark, or glueball, which possess properties that are forbidden for conventional hadrons. The states that do not fit into the ordinary $q\bar{q}$ or $qqq$ scheme in the quark model are referred to as exotic states.

The experimental discovery of exotic states began in 2003 with the observation of the $X(3872)$~\cite{3872-1}. This new state did not fit any ordinary $c\bar{c}$ quarkonia in the quark model. After that, the $X(3872)$ was observed in multiple decay modes and confirmed by various experiments~\cite{3872-2,3872-3,3872-4}. Many different theoretical interpretations of this state have been proposed, such as meson molecule, tetraquark, and conventional bound state~\cite{3872-theory-1,3872-theory-2,3872-theory-3,3872-theory-4,3872-theory-5}.
During the past two decades, there has been considerable world-wide activity in exotic state research using various processes, such as $e^+e^-$ annihilation (e.g., at $\tau$-charm facilities and B-factories), hadron collisions (e.g., at the Tevatron and the LHC), or photo- and leptoproduction (e.g., at the SPS, HERA or at Jefferson Lab),
and many exotic state candidates were observed~\cite{xyz-yuan,xyz-shen}. 

In searches for exotic states, a clear feature that helps distinguish exotic from ordinary hadrons would be a nonzero electric charge in a state which contains a heavy quark-antiquark pair of the same flavor. Such a state must contain at least one more quark-antiquark pair, and is thus not a conventional quark-antiquark meson.
Furthermore, a state with a pair of two identical heavy flavor quarks (for example, $cc$), has even more pronounced features as an exotic state.
Very recently, the LHCb experiment announced observation of an open-double-charm state $T_{cc}^+$ in the $D^{0}D^{0}\pi^{+}$ mass spectrum near threshold~\cite{tcc-lhcb-1,tcc-lhcb-2}.
It contains two charm quarks and two light quarks, thus it is a clear evidence for an exotic state. On the theoretical side, in addition to tetraquark models based on a heavy quark pair and two light quarks, the double-heavy tetraquark states are studied using QCD sum rules~\cite{the_qcd}, quark models~\cite{the_qm1,the_qm2}, and lattice QCD computations~\cite{the_lattqcd}. Besides, a QCD-inspired chiral quark model gives a prediction on the tetraquark states denoted as $\ccss$ with $+2$ electric charge in spin-parity channels $J^{P} = 0^{+}$ and $2^{+}$, which are expected to be found in $\dsds$ and $\dstdst$ final states~\cite{the_qqss2}. The predicted masses and widths of those resonances are listed in Table~\ref{tab:theoritical_predict}. Among the three predicted resonances in $\dstdst$ final state, the narrowest one has the highest observable probability.

\begin{table}[h]
  \begin{center}
    \caption{\label{tab:theoritical_predict} Predicted masses and widths for the $\ccss$ resonances in $D_{s}^{+}D_{s}^{+}$ and $D_{s}^{*+} D_{s}^{*+}$ final states~\cite{the_qqss2}. }
    \renewcommand\arraystretch{1.3}
    \begin{tabular}{cccc}
      \hline
      \hline
      Mode & $~~IJ^{P}$~~ &    ~~Mass~~          & ~~Width~~ \\
           &              & ~~(MeV/$c^2$)~~      & ~~(MeV)~~ \\
      \hline
            $\ccss \to \dsds$     & 00$^{+}$ & 4902 & 3.54 \\
            $\ccss \to \dstdst$   & 02$^{+}$ & 4821 & 5.58 \\
               & 02$^{+}$ & 4846 & 10.68 \\
               & 02$^{+}$ & 4775 & 23.26 \\

      \hline
      \hline
      \end{tabular}
    \end{center}
  \end{table}

In this paper, we present a search for double-heavy tetraquark candidates using the $D_{s}^{+}D_{s}^{+}$ and $D_{s}^{*+} D_{s}^{*+}$ final states in $\onetwos$ inclusive decays, and $\EE \to D_{s}^{+}D_{s}^{+}(D_{s}^{*+}D_{s}^{*+}) + anything$ processes at $\sqrt{s}$ = 10.52, 10.58, and 10.867~GeV. The $D_{s}^{*+}$ candidates are reconstructed in decays to $D_s^+\gamma$, while the $D_s^+$ candidates are reconstructed in the $D_s^+\to \phi(\to K^{+}K^{-})\pip$ and $\bar{K}^{*}(892)^{0}(\to K^{-}\pip)K^{+}$ decays. Inclusion of charged-conjugate modes is implicitly assumed throughout this analysis.

\section{\boldmath The data sample and the belle detector}
The data samples used in this analysis include: a 5.74 fb$^{-1}$ data sample collected
at the $\ones$ peak (102 million $\ones$ events); a 24.7 fb$^{-1}$ data sample collected
at the $\twos$ peak (158 million $\twos$ events); an 89.5 fb$^{-1}$ data sample collected at $\sqrt{s}$ = 10.52 GeV; a 711 fb$^{-1}$ data sample collected at $\sqrt{s}$ = 10.58 GeV, and a
121.4 fb$^{-1}$ data sample collected at $\sqrt{s}$ = 10.867 GeV, where $s$ is the
center-of-mass energy squared.
All the data were collected with the Belle detector, which is described in detail in Ref.~\cite{detector}, operating at the KEKB asymmetric-energy $e^+e^-$
collider~\cite{collider}.
It is a large-solid-angle magnetic spectrometer consisting of a silicon vertex detector,
a 50-layer central drift chamber (CDC), an array of aerogel threshold Cherenkov counters (ACC),
a barrel-like arrangement of time-of-flight scintillation counters (TOF), and an electromagnetic
calorimeter comprising CsI(Tl) crystals (ECL) located inside a superconducting solenoid coil that
provides a $1.5~\hbox{T}$ magnetic field. An iron flux return comprising resistive plate chambers
placed outside the coil was instrumented to detect $K^{0}_{L}$ mesons and to identify muons.

Monte Carlo (MC) signal events are generated with {\sc EvtGen}~\cite{evtgen} and processed through a full simulation of the Belle detector based on {\sc GEANT3}~\cite{geant}.
Initial-state radiation (ISR) is taken into account assuming that the
cross sections follow a $1/s$ dependence in $\EE \to \ccss + anything$ reactions.
The processes $\onetwos \to D_{s}^{+}D_{s}^{+}(D_{s}^{*+}D_{s}^{*+}) + anything$ and $\EE \to D_{s}^{+}D_{s}^{+}(D_{s}^{*+}D_{s}^{*+}) + anything$ at $\sqrt{s}$ = 10.52~GeV, 10.58~GeV, and 10.867~GeV are taken into account, where the $D_{s}^{*+}$ decays into $D_{s}^{+} \gamma$ using a $P$-wave model, and the $D_{s}^{+}$ decays to $\KK\pi^{+}$ final states using a Dalitz plot decay model of Ref.~\cite{cleo-dalitz}. The mass of $\ccss$ is chosen in the interval from 4882~MeV/$c^{2}$ to 4922~MeV/$c^{2}$ (4801~MeV/$c^{2}$ to 4841~MeV/$c^{2}$)
in steps of 5~MeV/$c^{2}$, with a width varying from 0.54~MeV to 6.54~MeV (2.58~MeV to 8.58~MeV) in steps of 1~MeV for $\ccss \to D_{s}^{+}D_{s}^{+}$ ($D_{s}^{*+}D_{s}^{*+}$).
Inclusive MC samples of $\Upsilon(1S,2S)$ decays, $\fours \to B^{+}B^{-}/B^{0}\bar{B}^{0}$, $\fives \to B_{s}^{(*)} \bar{B}_{s}^{(*)}$, and $\EE \to q\bar{q}$ $(q=u, d, s, c)$ at $\sqrt{s}$ = 10.52~GeV, 10.58~GeV, and 10.867~GeV corresponding to four times the integrated luminosity of data are used to study possible peaking backgrounds.

\section{\boldmath Common Event selection criteria}
For reconstructed charged tracks,
the impact parameters perpendicular to and along the beam direction with respect to the interaction point (IP)
are required to be less than 0.2~cm and 1.5~cm, respectively, and the transverse momentum in the
laboratory frame is required to be larger than 0.1~GeV/$c$.
For the particle identification (PID) of a well-reconstructed charged track,
information from different detector subsystems, including specific ionization in the CDC,
time measurement in the TOF, and the response of the ACC, is combined to form a likelihood
${\mathcal L}_i$~\cite{pidcode} for particle species $i$, where $i$ =  $\pi$ or $K$.
Tracks with $R_{K}=\mathcal{L}_{\textrm{K}}/(\mathcal{L}_K+\mathcal{L}_\pi)<0.4$ are identified as
pions with an efficiency of 96\%, while 5\% of kaons are misidentified as pions; tracks
with $R_{K}>0.6$ are identified as kaons with an efficiency of 95\%, while 4\% of pions are
misidentified as kaons.

An ECL cluster is taken as a photon candidate if it does not match the extrapolation of any charged tracks.
The energy of the photon candidate from the $D_{s}^{*+}$ decay is required to be greater than 50 MeV.
For $D_{s}^{+}$ candidates, vertex and mass-constrained fits are performed, and then $\chi^{2}_{\textrm{vertex}}/n.d.f < 20$ is required ($> 97\%$ selection efficiency according to MC simulation).
For $D_{s}^{*+}$ candidates, a mass-constrained fit is performed to improve its momentum resolution.
The best $D_{s}^{*+}$ candidate with $\chi^{2}$ of $D_{s}^{*+}$ mass-constrained fit for each $D_{s}^{+}$ candidate is kept to suppress the combinational background.

\begin{figure*}[htbp]
  \begin{center}
    \includegraphics[height=3.75cm,width=5cm]{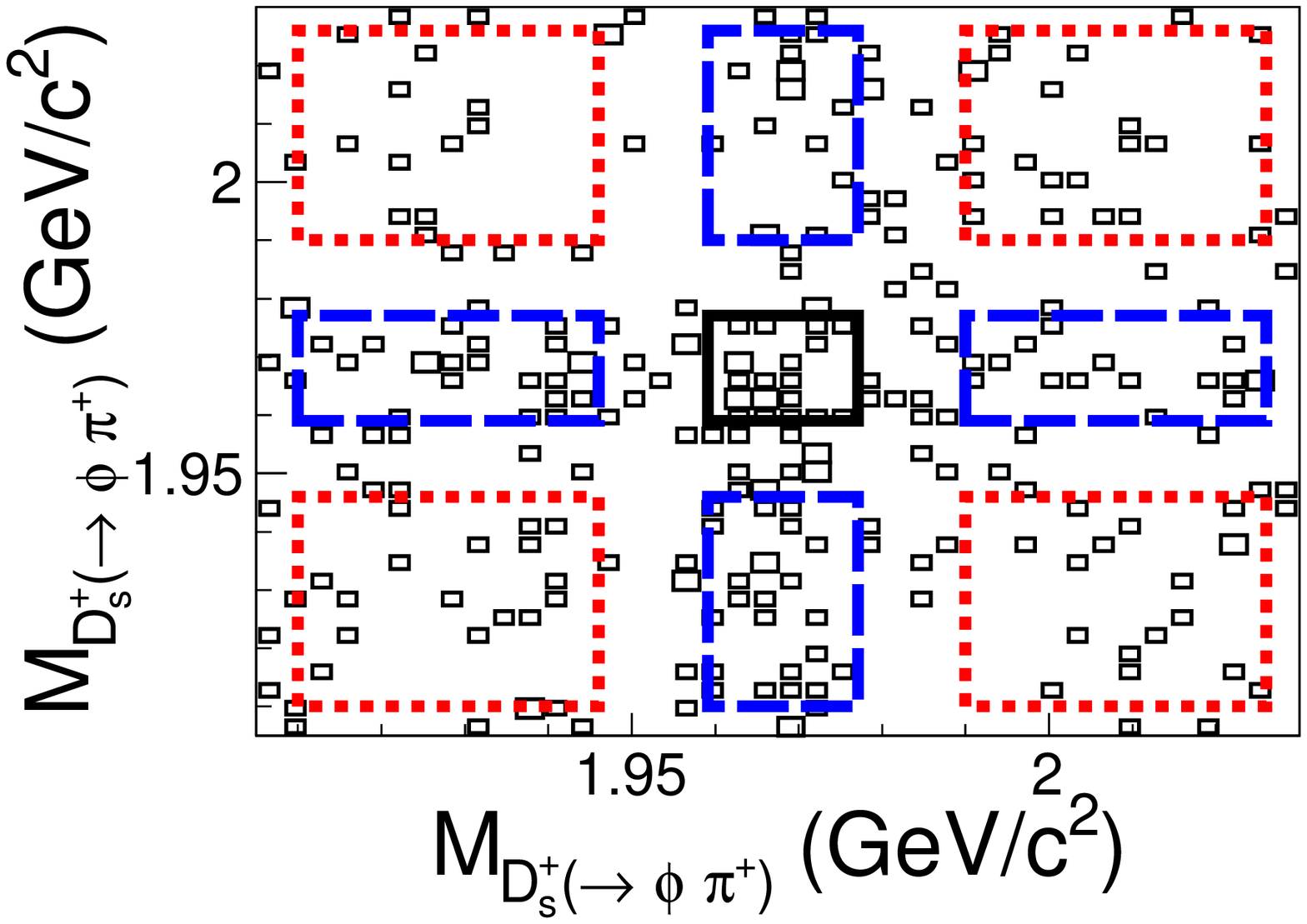}
    \includegraphics[height=3.75cm,width=5cm]{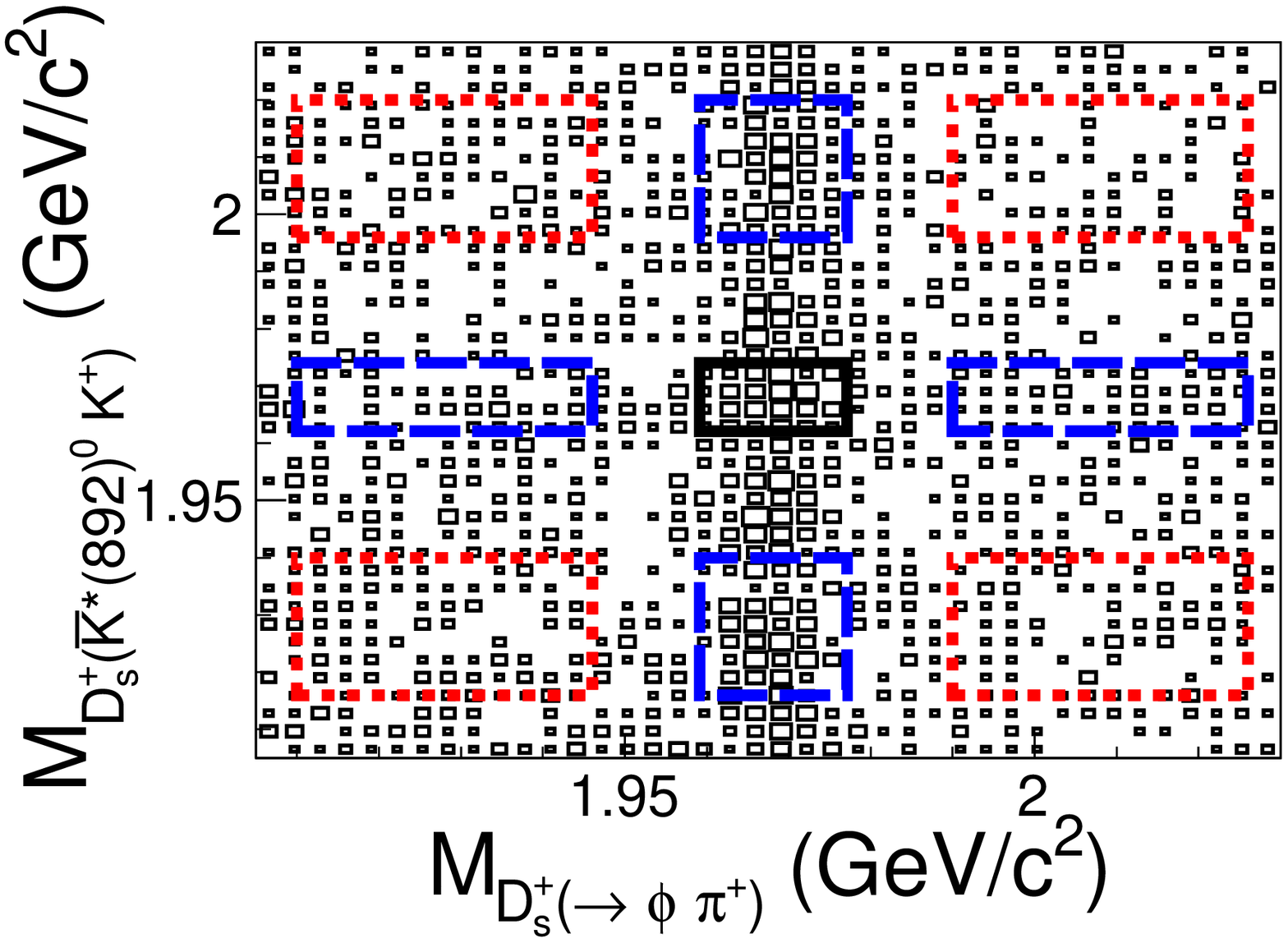}
    \includegraphics[height=3.75cm,width=5cm]{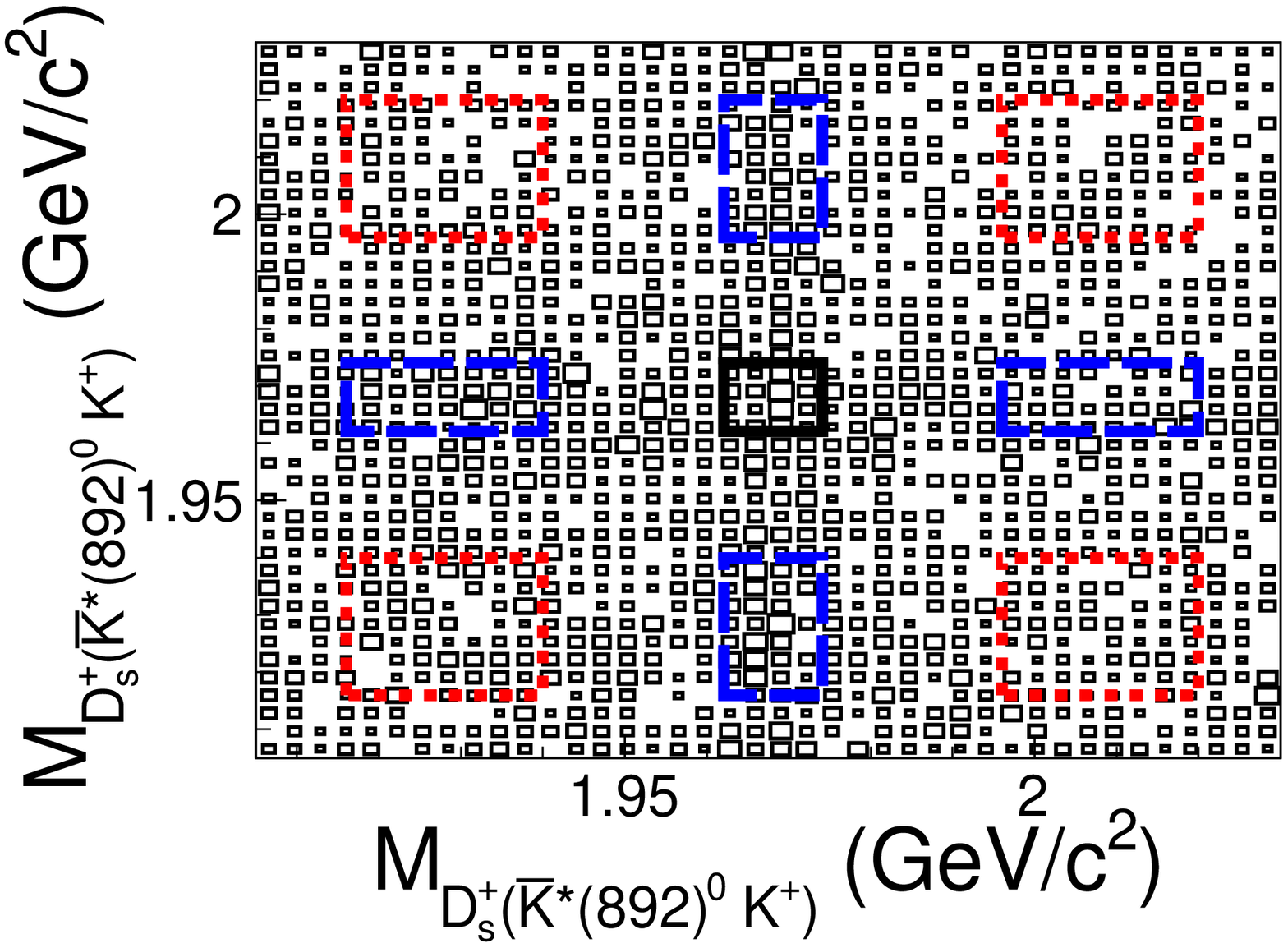}\\
    \includegraphics[height=3.75cm,width=5cm]{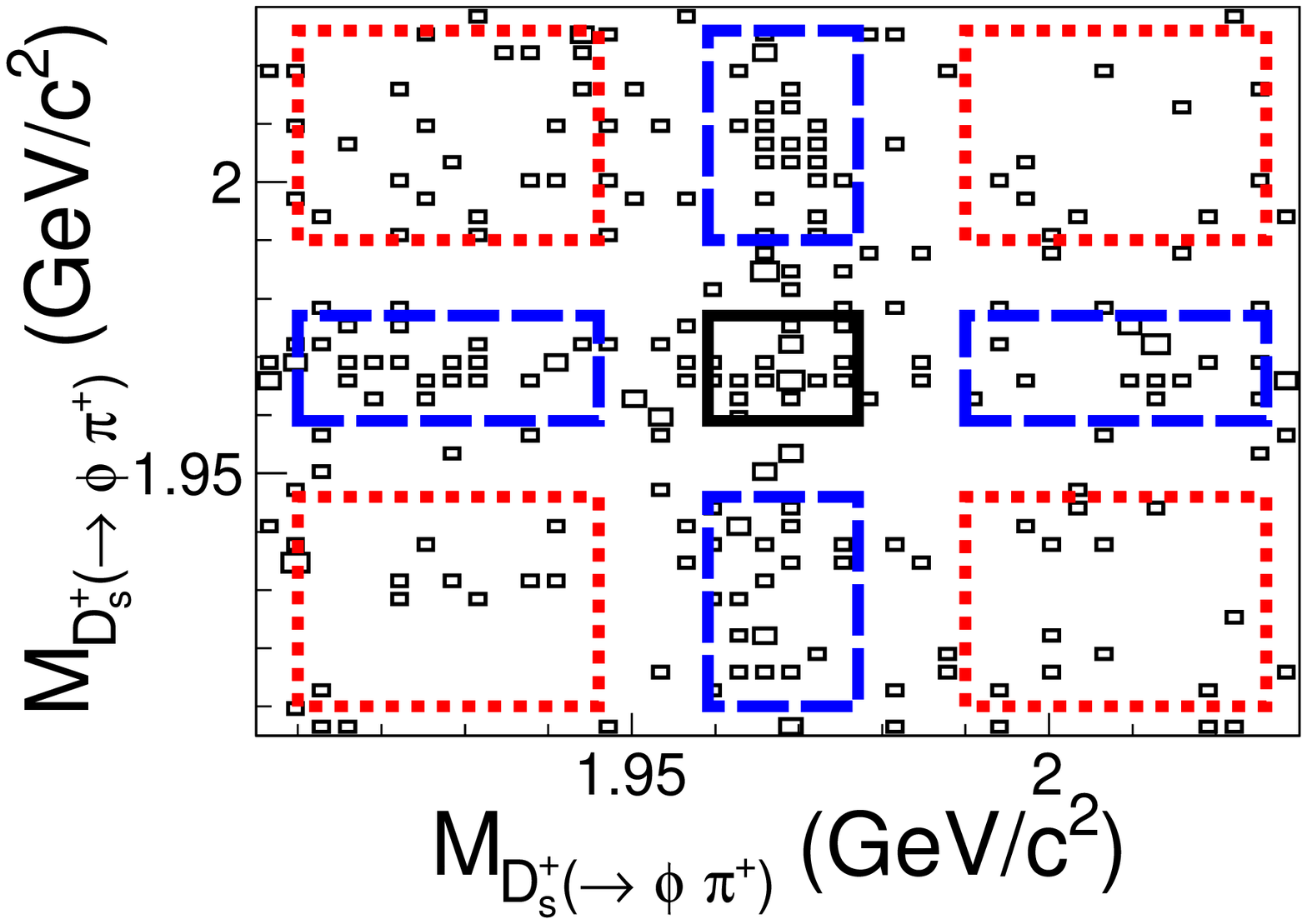}
    \includegraphics[height=3.75cm,width=5cm]{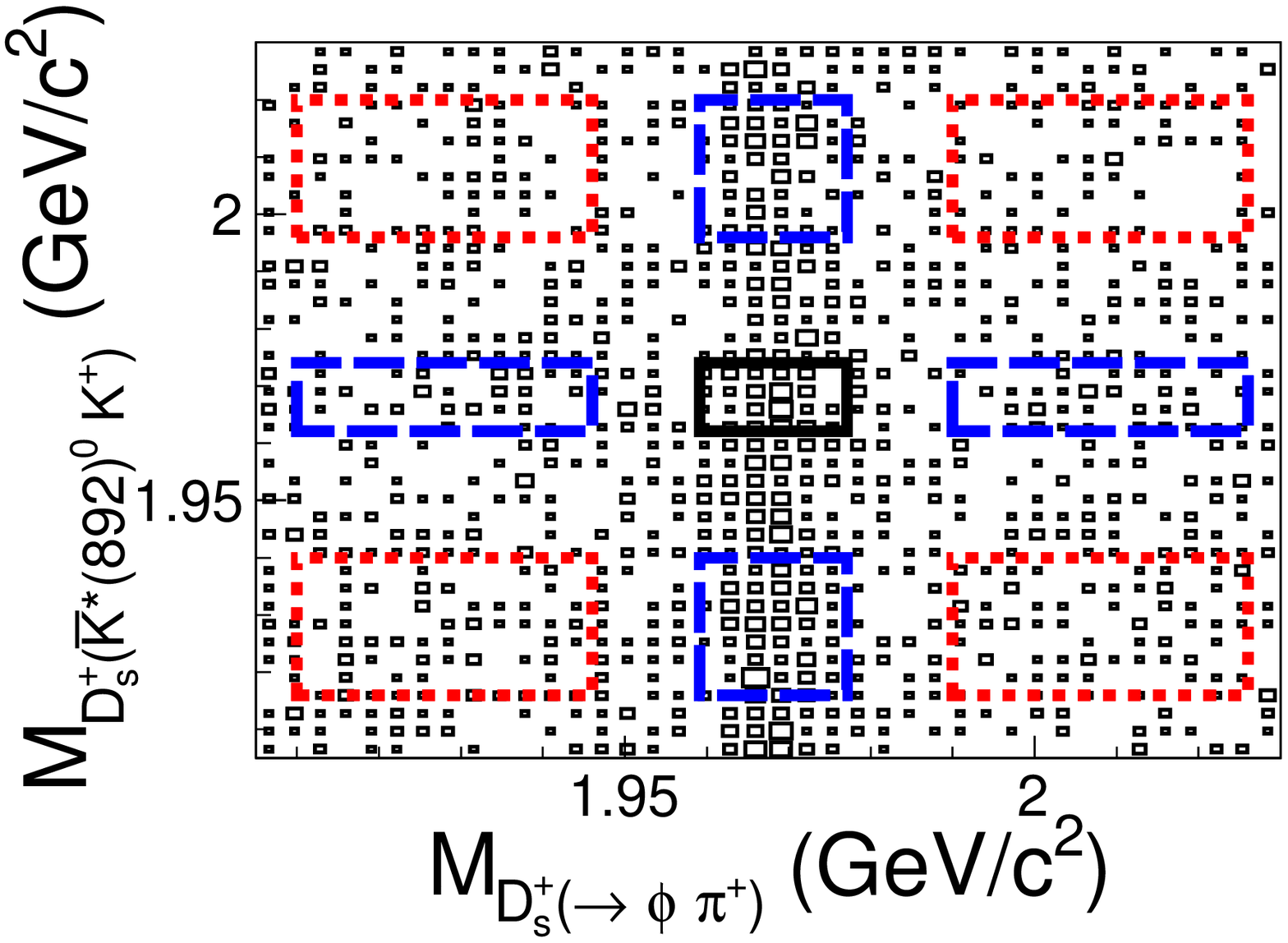}
    \includegraphics[height=3.75cm,width=5cm]{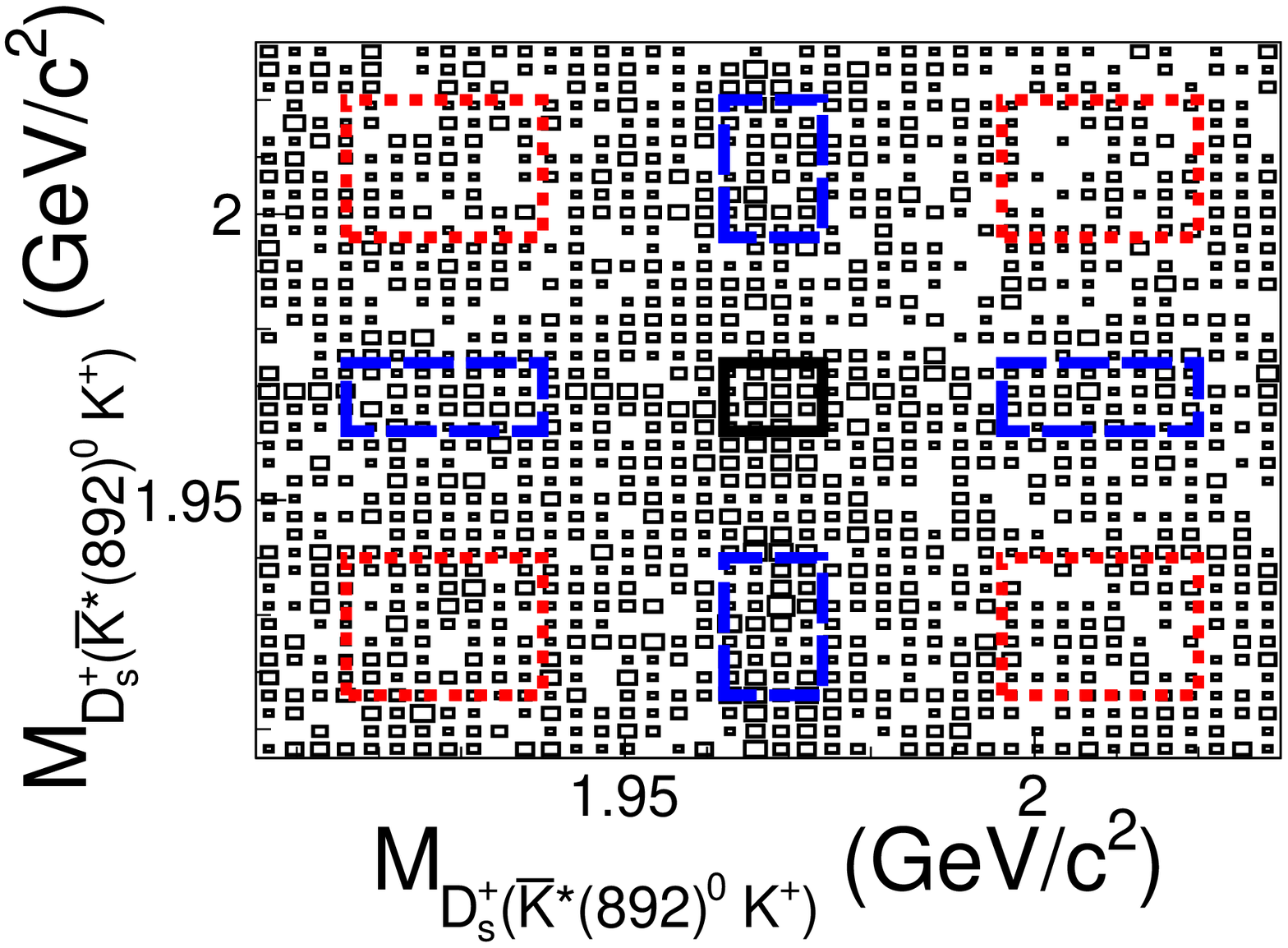}
    \caption{The top (bottom) plots show the distribution of $M_{D_{s}^{+}}$ vs $M_{D_{s}^{+}}$ from the selected $\EE \to \ccss \to D_{s}^{+}D_{s}^{+}~ (D_{s}^{*+}D_{s}^{*+}) + anything$ candidates from data at $\sqrt{s}$ = 10.58~GeV, where the $D_{s}^{+}$ is reconstructed from $\phi \pip$ or $\bar{K}^{*}(892)^{0} K^{+}$. The central solid boxes define the signal regions, and the red dash-dotted and blue dashed boxes show the $M_{D_{s}^{+}}$ sideband regions described in the text.}\label{DD_mass_data_2D}
  \end{center}
\end{figure*}

The signal mass windows for $\bar{K}^{*}(892)^{0}$, $\phi$, $D_s^+$, and $D_{s}^{*+}$ candidates have been optimized by maximizing the Punzi parameter $S/(3/2+\sqrt{B})$~\cite{fom},
where $S$ is the number of selected events in the simulated signal process by fitting the $\ccss$ invariant-mass spectrum.
$B$ is the number of selected events obtained from the normalized $M_{D_{s}^{+}D_{s}^{+}}$ sidebands in inclusive MC samples.
The optimized mass window requirements are $|M_{K^{+}K^{-}} - m_{\phi}| < 8$ MeV/$c^{2}$, $|M_{\phi \pip} - m_{D_{s}^{+}}| < 7$ MeV/$c^{2}$, $|M_{K^{-}\pip} - m_{\bar{K}^{*}(892)^{0}}| < 50$ MeV/$c^{2}$, $|M_{\bar{K}^{*}(892)^{0} K^{+}} - m_{D_{s}^{+}}| < 7$ MeV/$c^{2}$, and $|M_{\gamma D_{s}^{+}} - m_{D_{s}^{*+}}| < 14$ MeV/$c^{2}$, where $m_{\phi}$, $m_{\bar{K}^{*}(892)^{0}}$, $m_{{D}_{s}^{+}}$, and $m_{{D}_{s}^{*+}}$ are the nominal masses of $\phi$, $\bar{K}^{*}(892)^{0}$, ${D}_{s}^{+}$, and ${D}_{s}^{*+}$~\cite{PDG}.
There are no multiple candidates after processing all selections in both $D_{s}^{+}D_{s}^{+}$ and $D_{s}^{*+}D_{s}^{*+}$ cases.
Figure~\ref{DD_mass_data_2D} shows the scatter plots of $D_{s}^{+}$ versus $D_{s}^{+}$ invariant masses from the selected
$\EE \to \ccss (\to D_{s}^{+}D_{s}^{+}(D_{s}^{*+}D_{s}^{*+})) + anything$ candidates from data at $\sqrt{s}$ = 10.58~GeV as an example. Here we define the two-dimensional $D_{s}^{+}D_{s}^{+}$ sidebands, and the normalized contribution from $D_{s}^{+}$ and $D_{s}^{+}$ sidebands is estimated using 25\% of the number of events
in the blue dashed line boxes and reduced by 6.25\% of the number of events in the red dotted line boxes.

\section{\boldmath Invariant-mass spectra}

\begin{figure*}[htbp]
  \begin{center}
    \includegraphics[height=3.75cm,width=5cm]{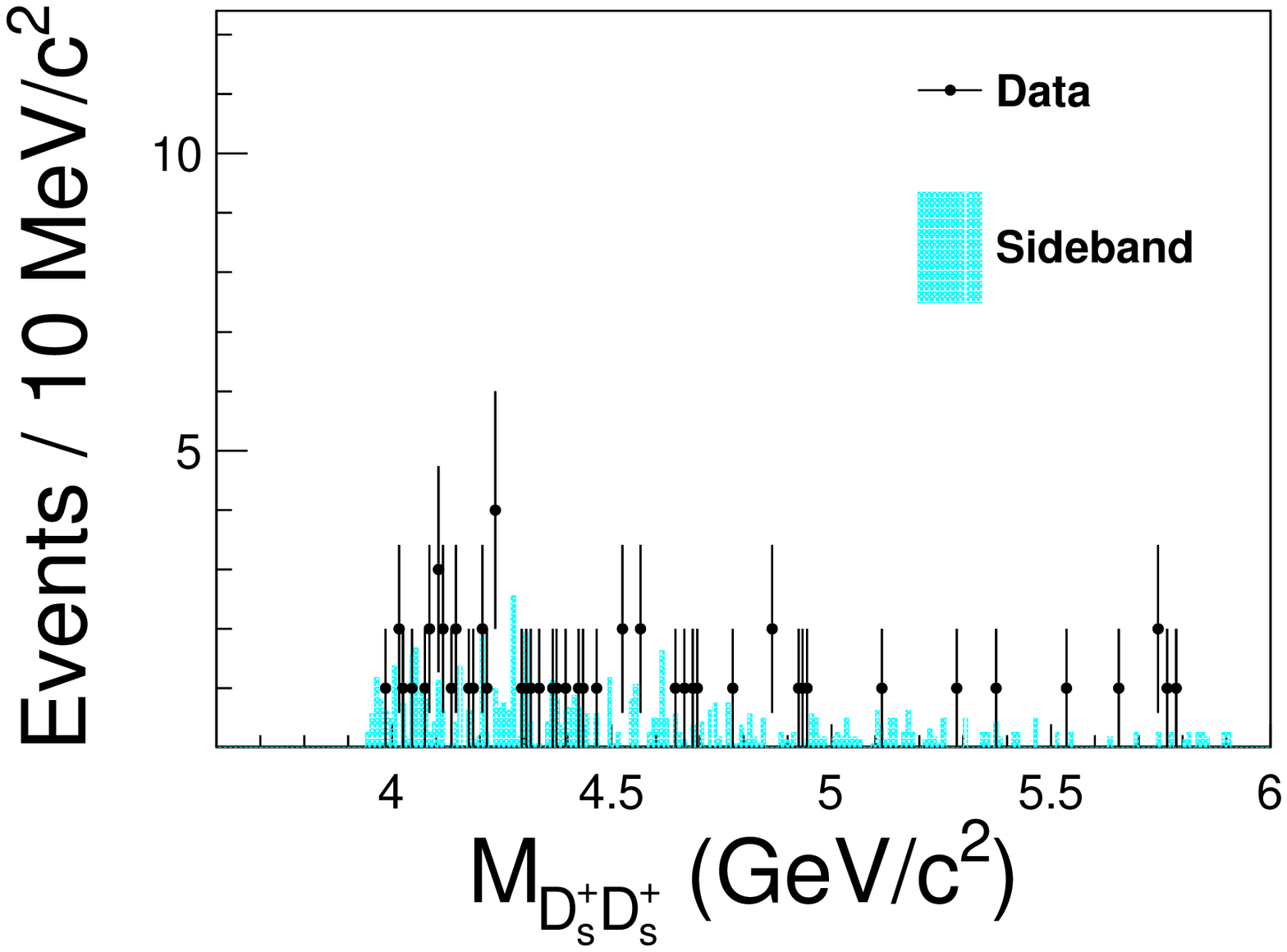}
    \includegraphics[height=3.75cm,width=5cm]{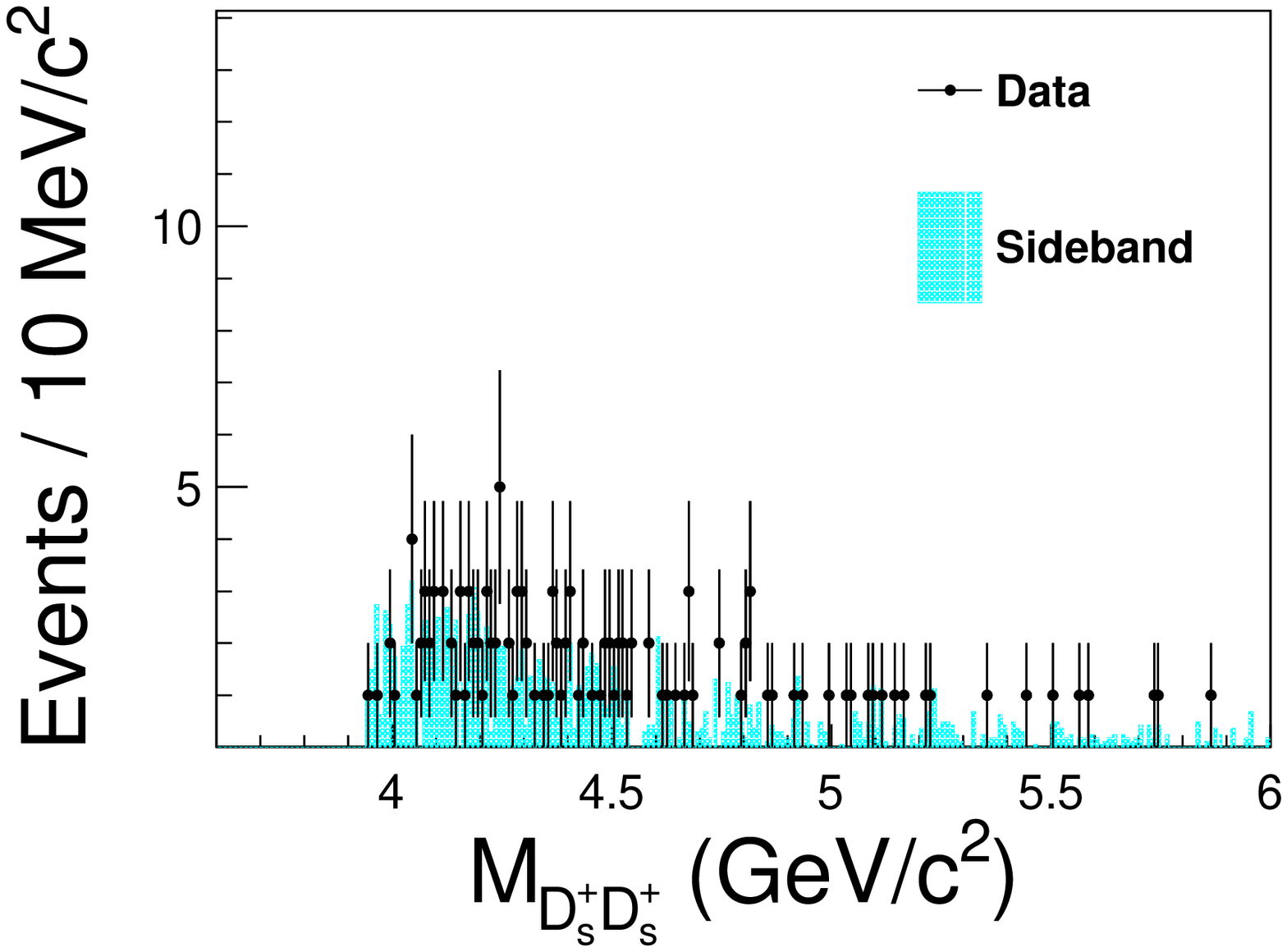}
    \put(-250,90){\bf (a)} \put(-105,90){\bf (b)}
\\
    \includegraphics[height=3.75cm,width=5cm]{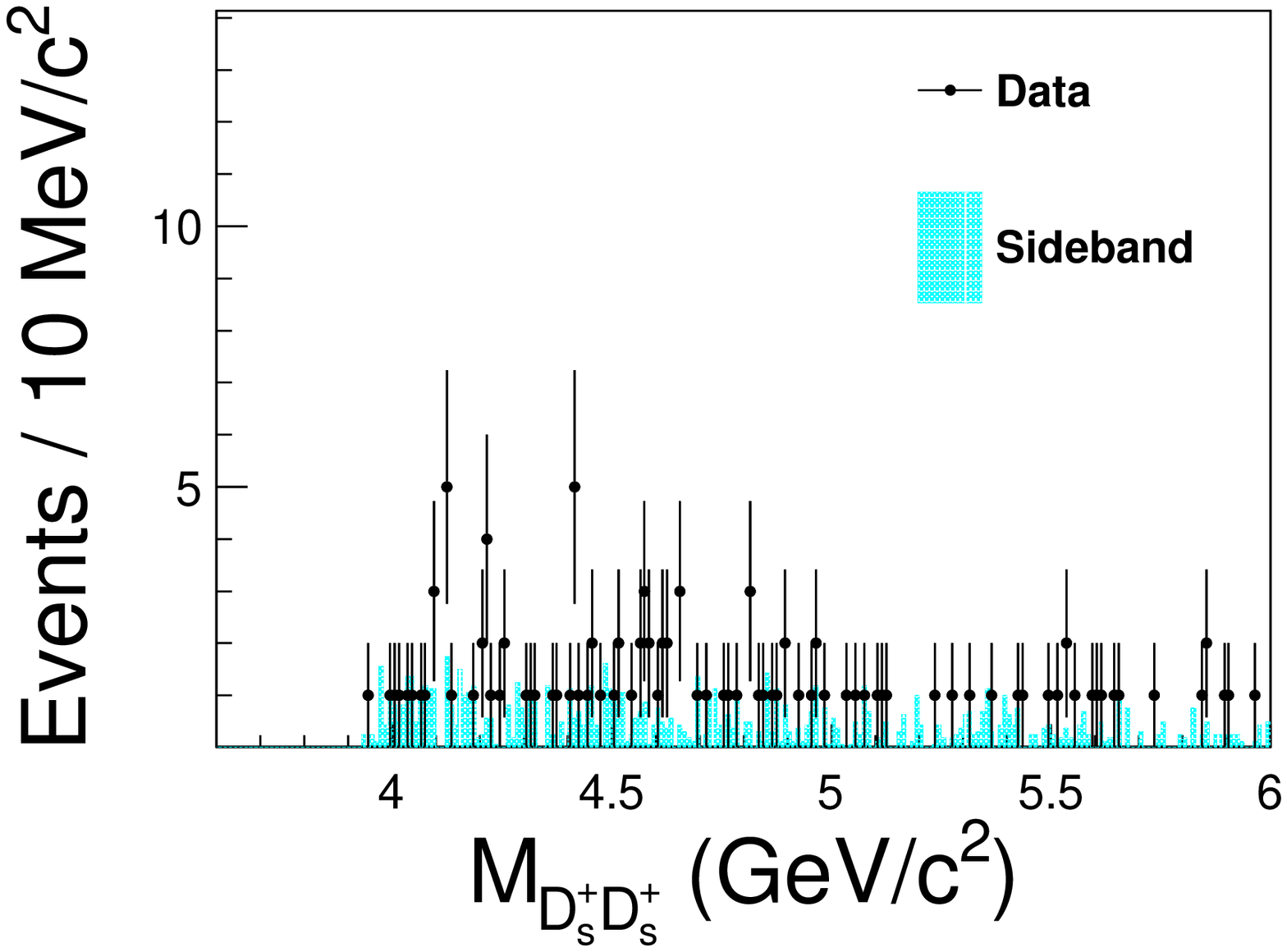}
    \includegraphics[height=3.75cm,width=5cm]{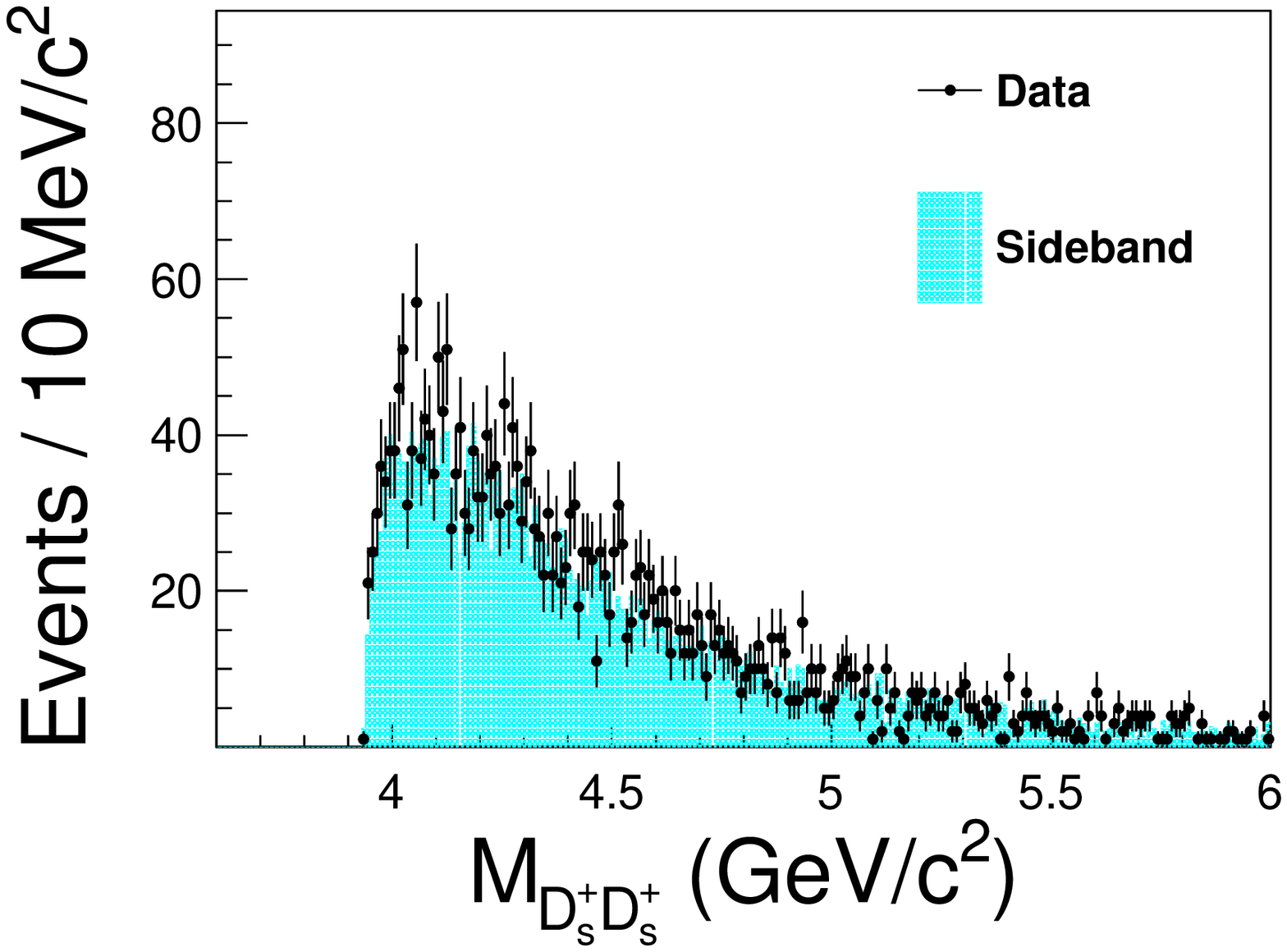}
    \includegraphics[height=3.75cm,width=5cm]{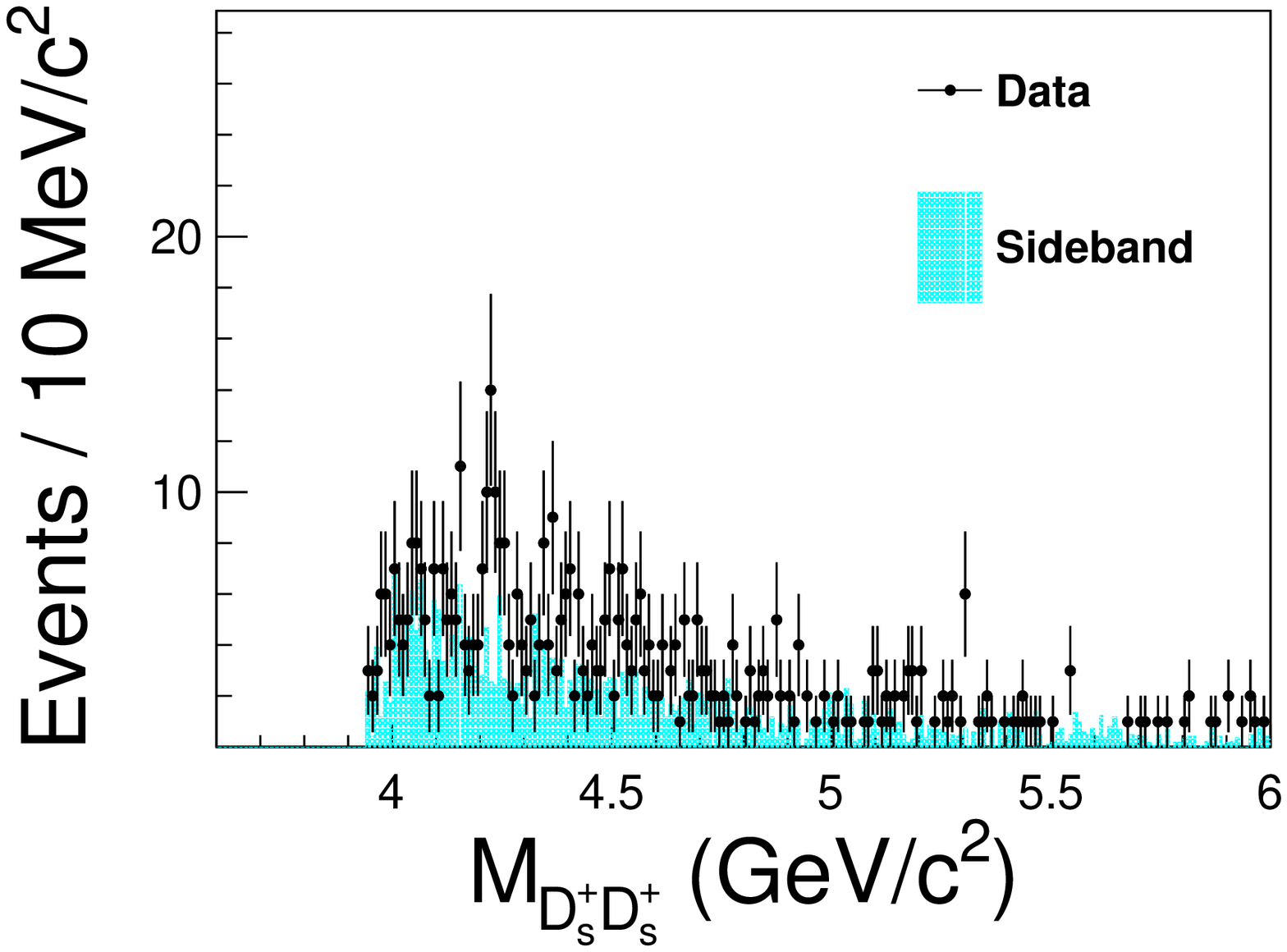}
    \put(-395,90){\bf (c)} \put(-250,90){\bf (d)} \put(-105,90){\bf (e)}
    \caption{Distributions of $M_{D_{s}^{+}D_{s}^{+}}$ from data for processes (a) $\ones \to \ccss (\to D_{s}^{+}D_{s}^{+}) + anything$, (b) $\twos \to \ccss (\to D_{s}^{+}D_{s}^{+}) + anything$, and $\EE \to \ccss (\to D_{s}^{+}D_{s}^{+}) + anything$ at (c) $\sqrt{s}$ = 10.52~GeV, (d) $\sqrt{s}$ = 10.58~GeV, (e) $\sqrt{s}$ = 10.867~GeV. The cyan shaded histograms are from the normalized $M_{D_{s}^{+}D_{s}^{+}}$ sideband events.}\label{DD_mass_data_all}
  \end{center}
\end{figure*}

\begin{figure*}[htbp]
  \begin{center}
    \includegraphics[height=3.75cm,width=5cm]{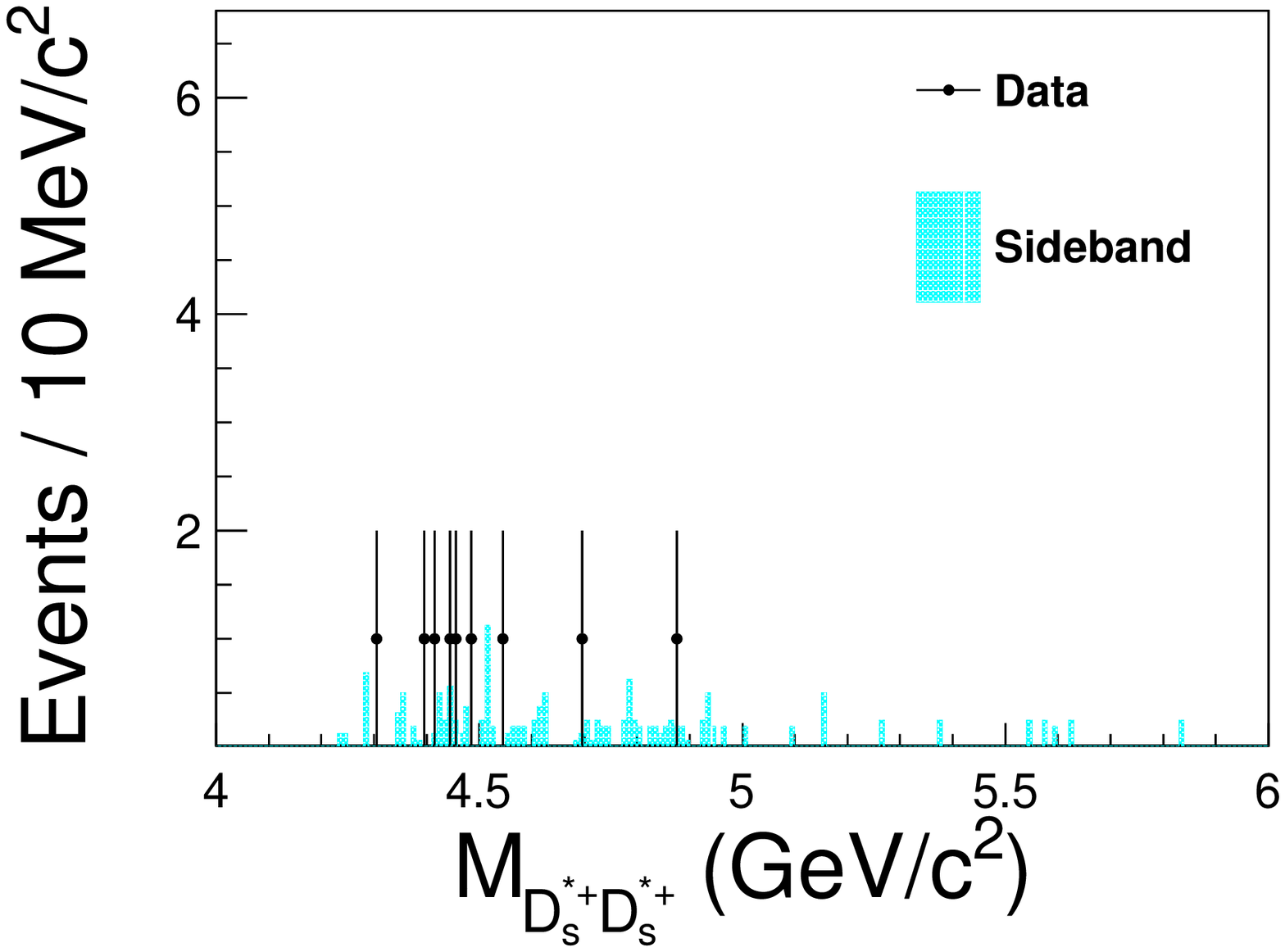}
    \includegraphics[height=3.75cm,width=5cm]{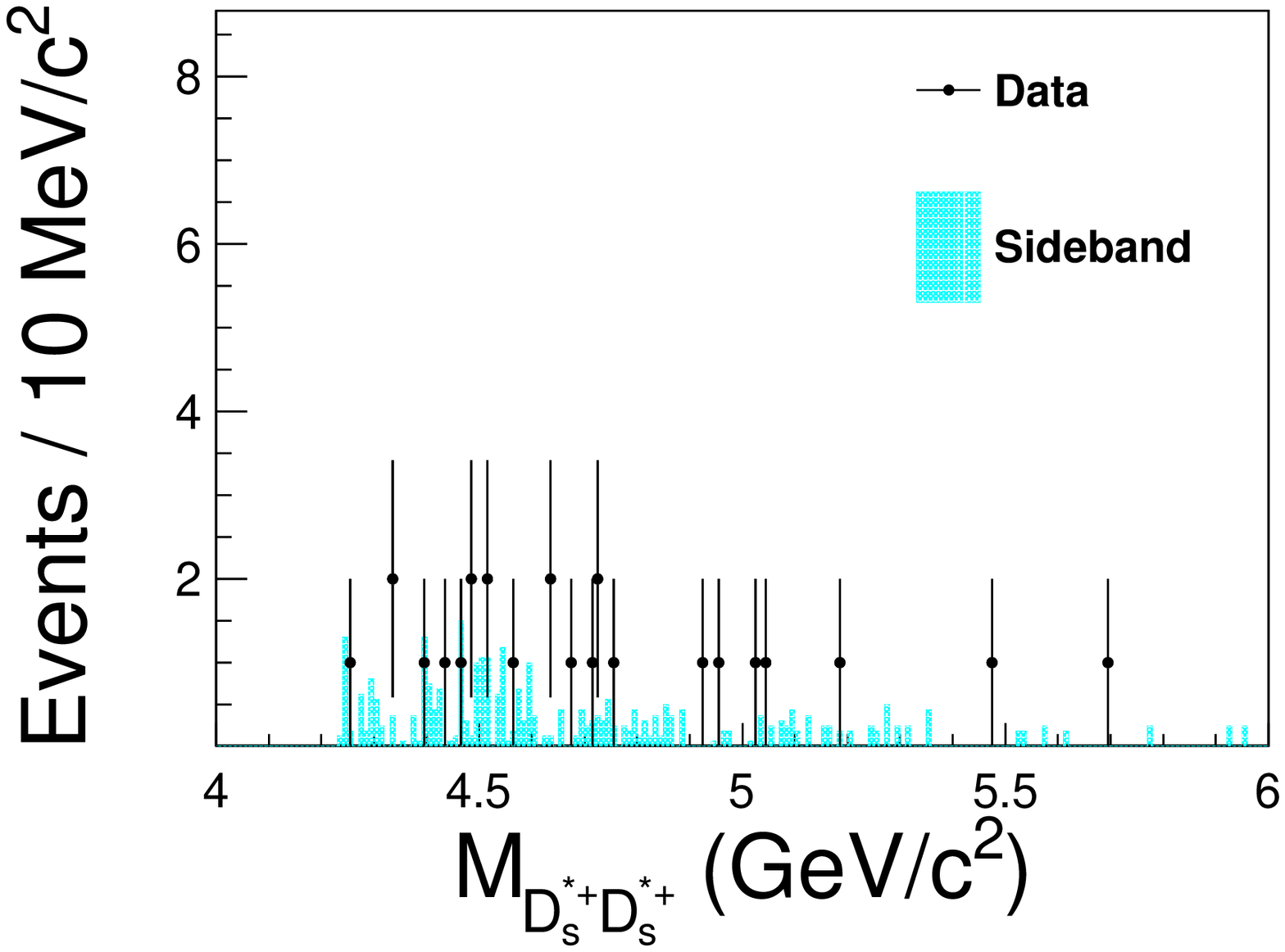}
    \put(-250,90){\bf (a)} \put(-105,90){\bf (b)}
\\
    \includegraphics[height=3.75cm,width=5cm]{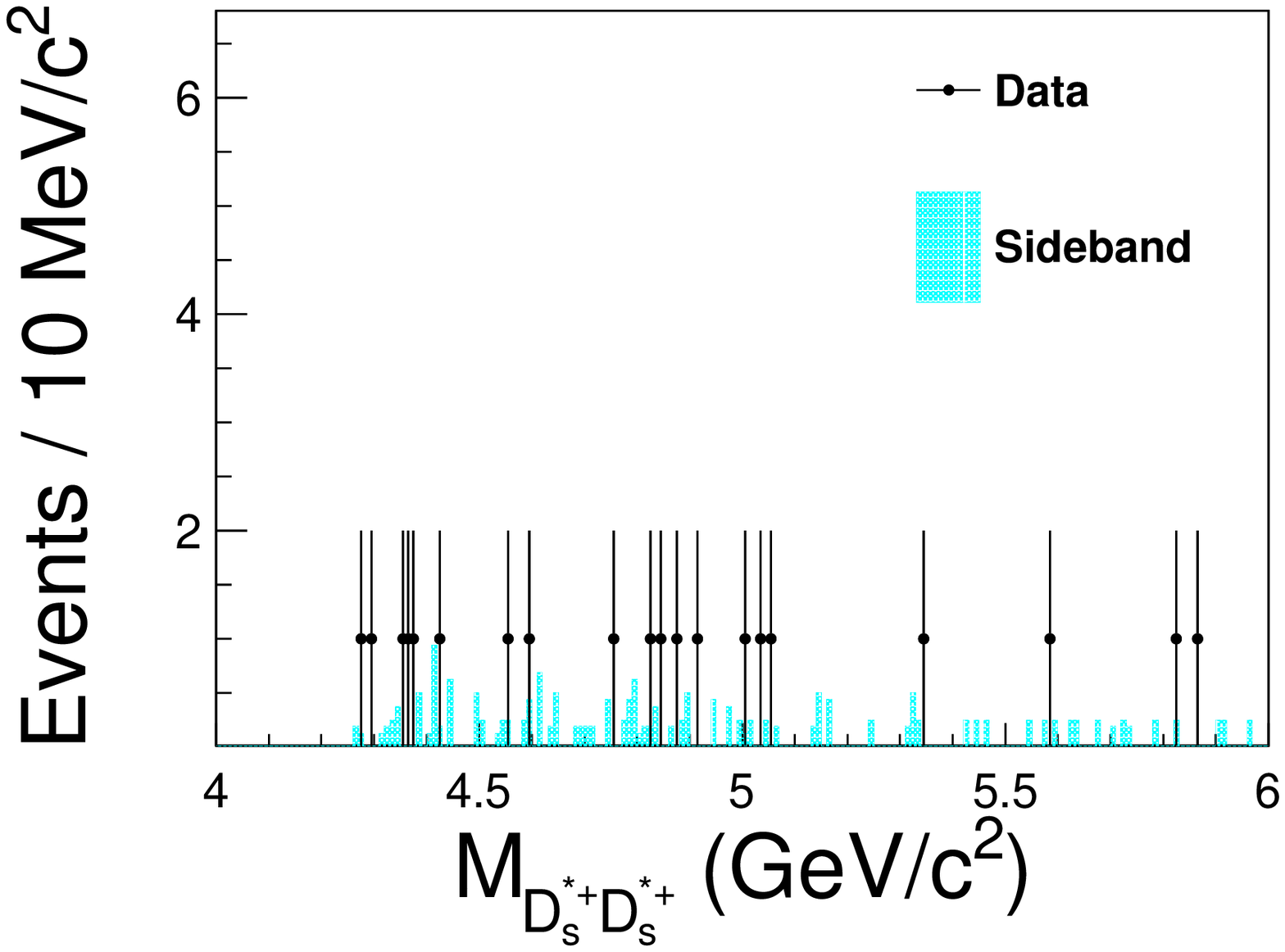}
    \includegraphics[height=3.75cm,width=5cm]{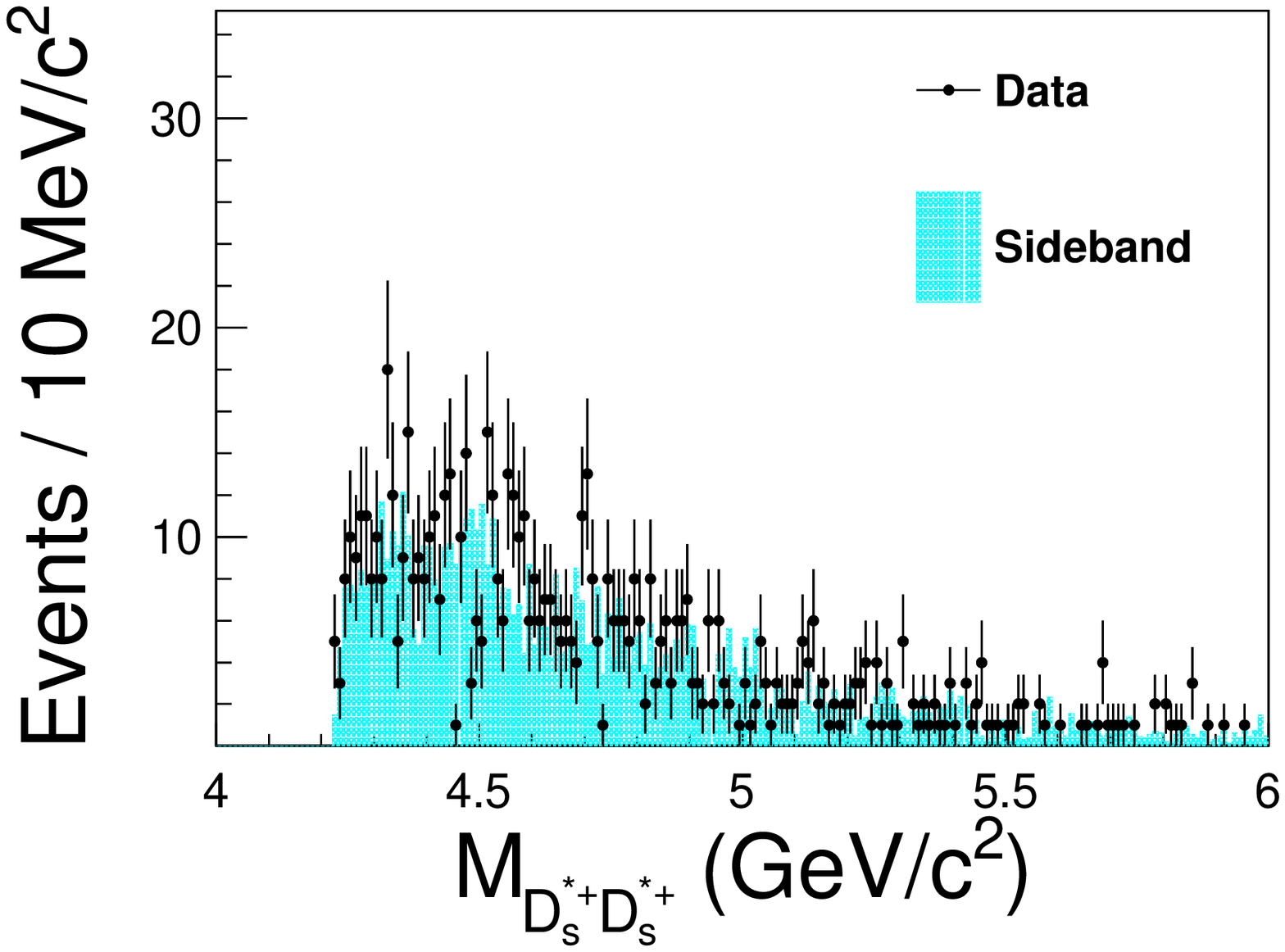}
    \includegraphics[height=3.75cm,width=5cm]{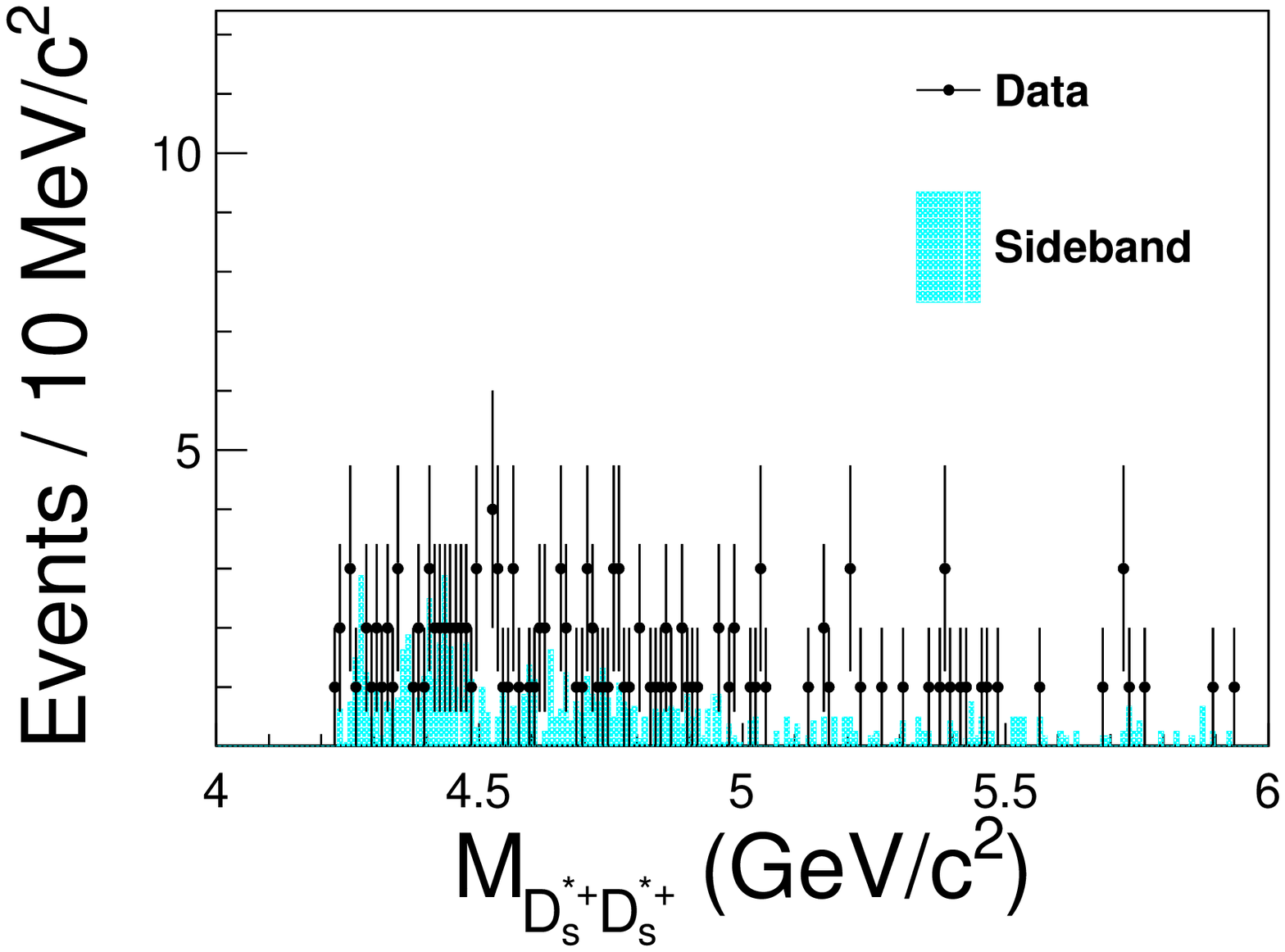}
    \put(-395,90){\bf (c)} \put(-250,90){\bf (d)} \put(-105,90){\bf (e)}
    \caption{Distributions of $M_{D_{s}^{*+}D_{s}^{*+}}$ from data for processes (a) $\ones \to \ccss (\to D_{s}^{*+}D_{s}^{*+}) + anything$, (b) $\twos \to \ccss (\to D_{s}^{*+}D_{s}^{*+}) + anything$, and $\EE \to \ccss (\to D_{s}^{*+}D_{s}^{*+}) + anything$ at (c) $\sqrt{s}$ = 10.52~GeV, (d) $\sqrt{s}$ = 10.58~GeV, (e) $\sqrt{s}$ = 10.867~GeV. The cyan shaded histograms are from the normalized $M_{D_{s}^{+}D_{s}^{+}}$ sideband events.}\label{DsDs_mass_data_all}
  \end{center}
\end{figure*}

\begin{figure*}[htbp]
  \begin{center}
    \includegraphics[height=3.75cm,width=5cm]{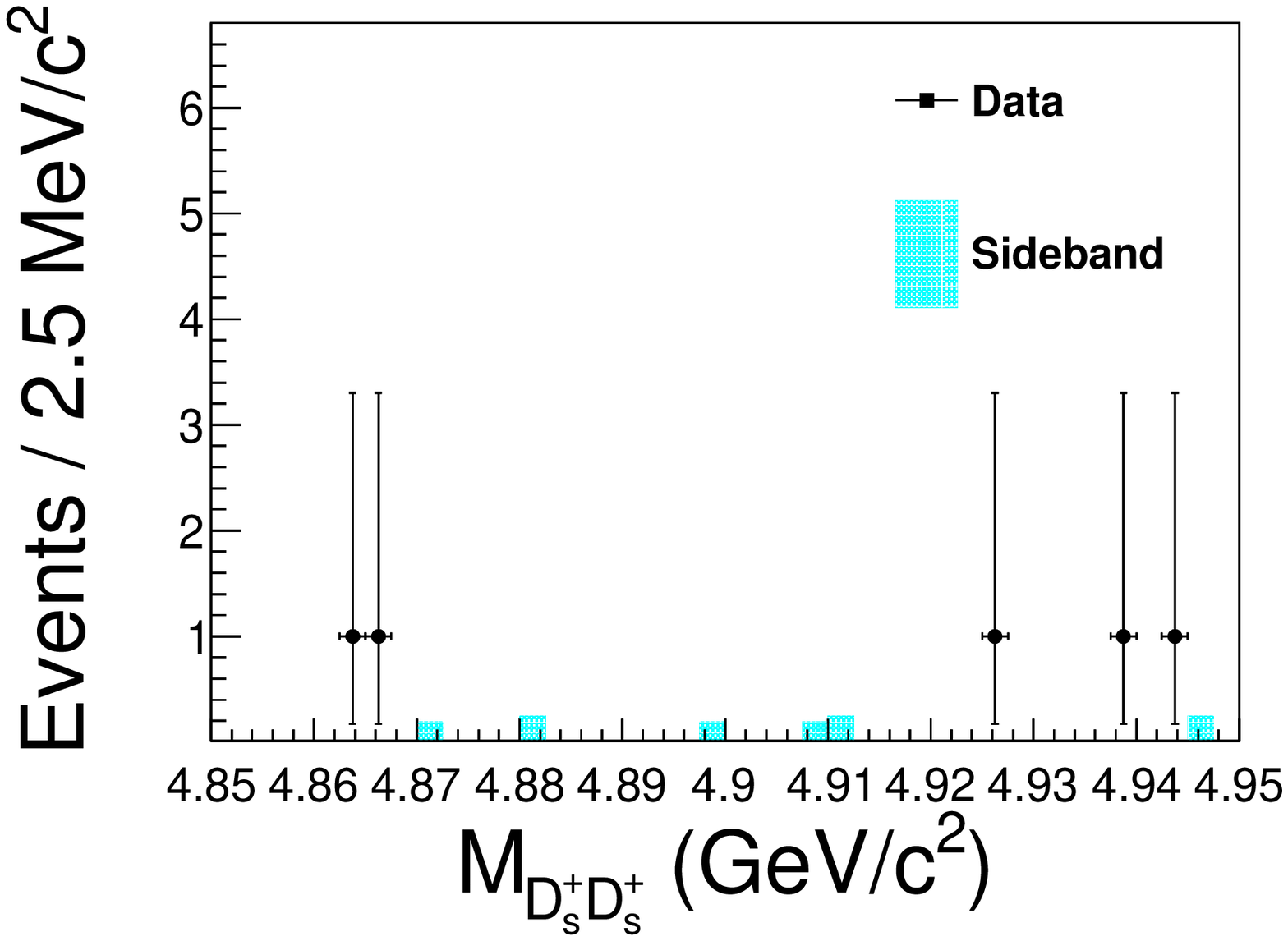}
    \includegraphics[height=3.75cm,width=5cm]{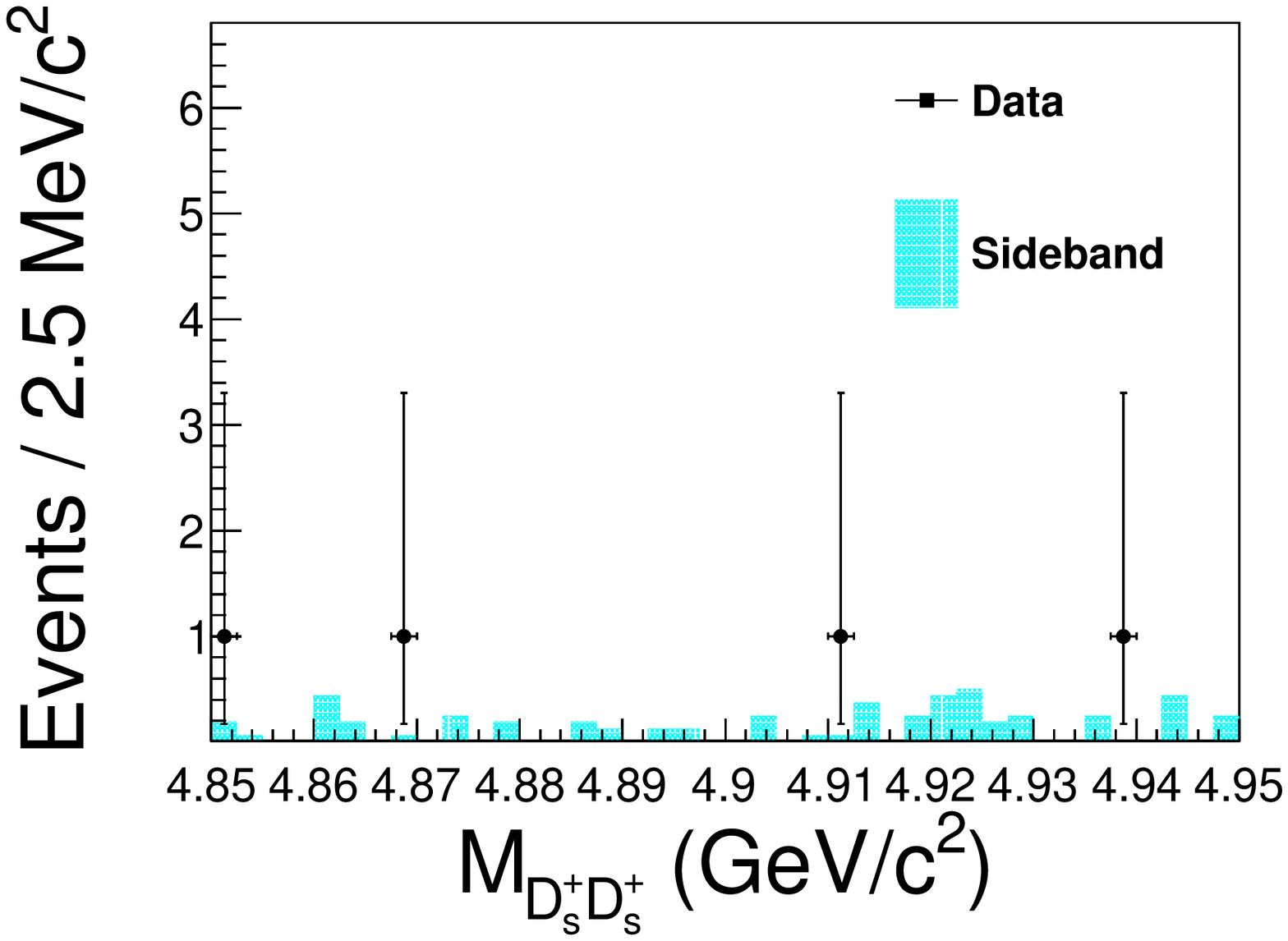}
    \put(-250,90){\bf (a)} \put(-105,90){\bf (b)}
\\
    \includegraphics[height=3.75cm,width=5cm]{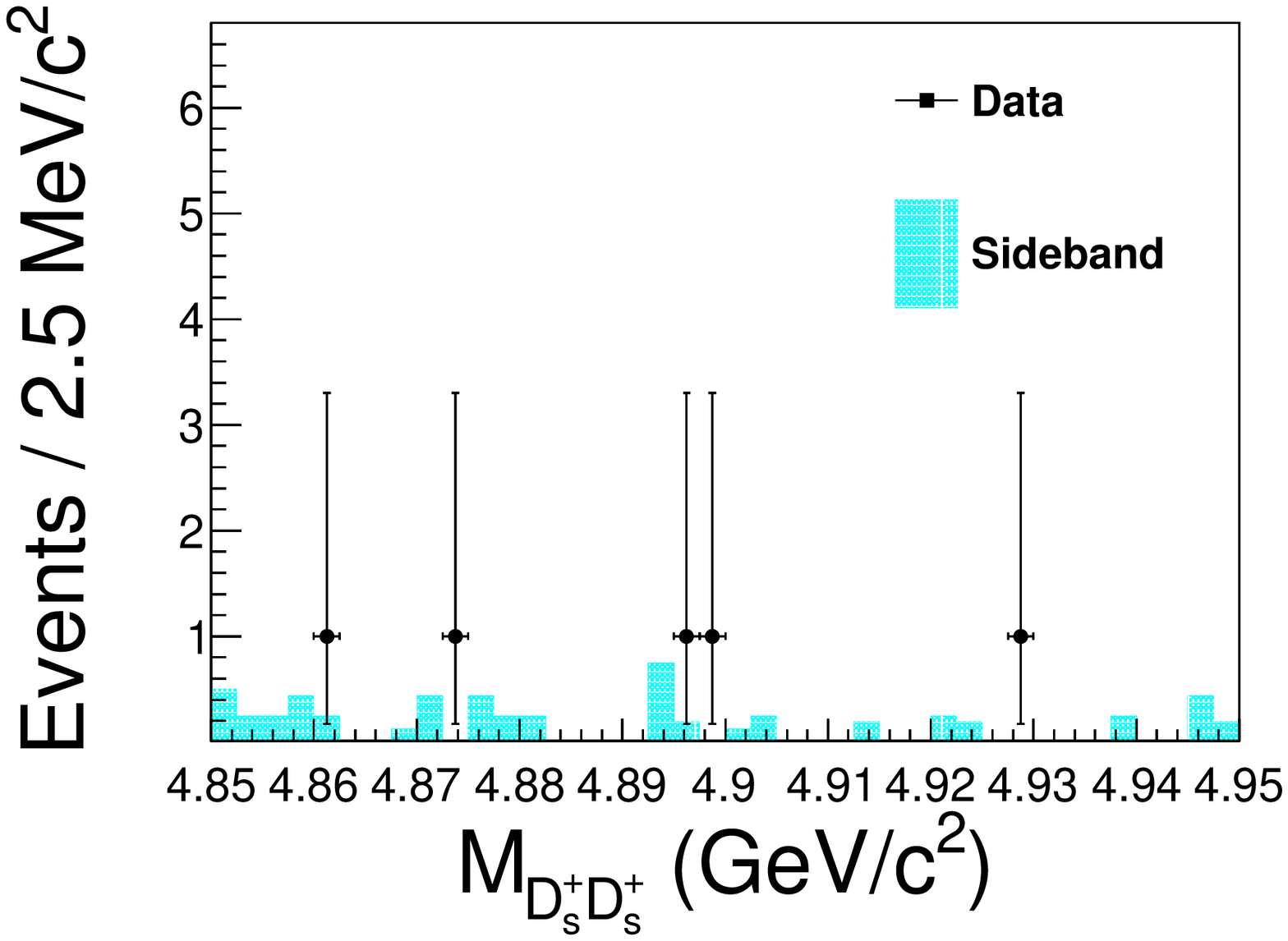}
    \includegraphics[height=3.75cm,width=5cm]{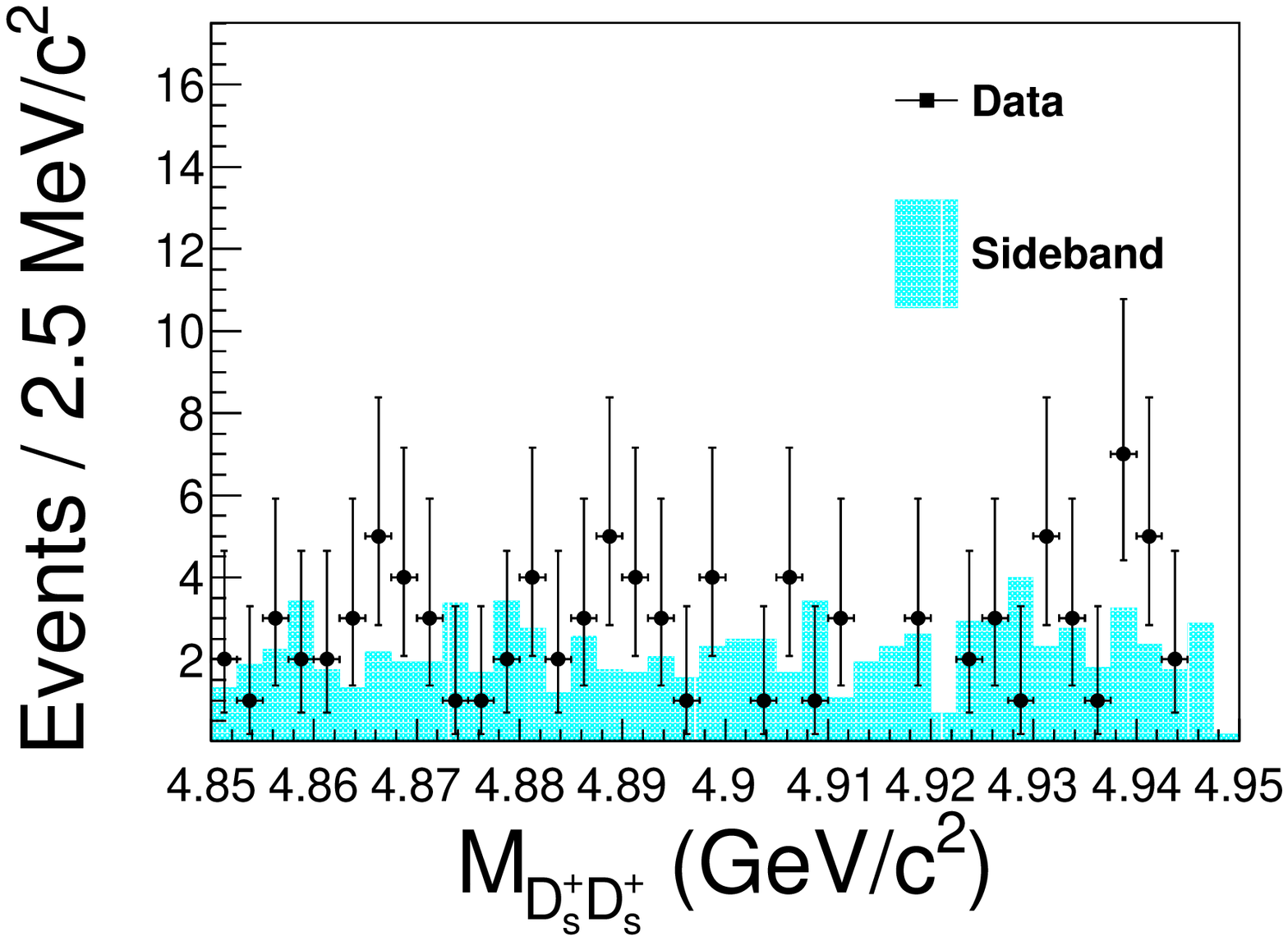}
    \includegraphics[height=3.75cm,width=5cm]{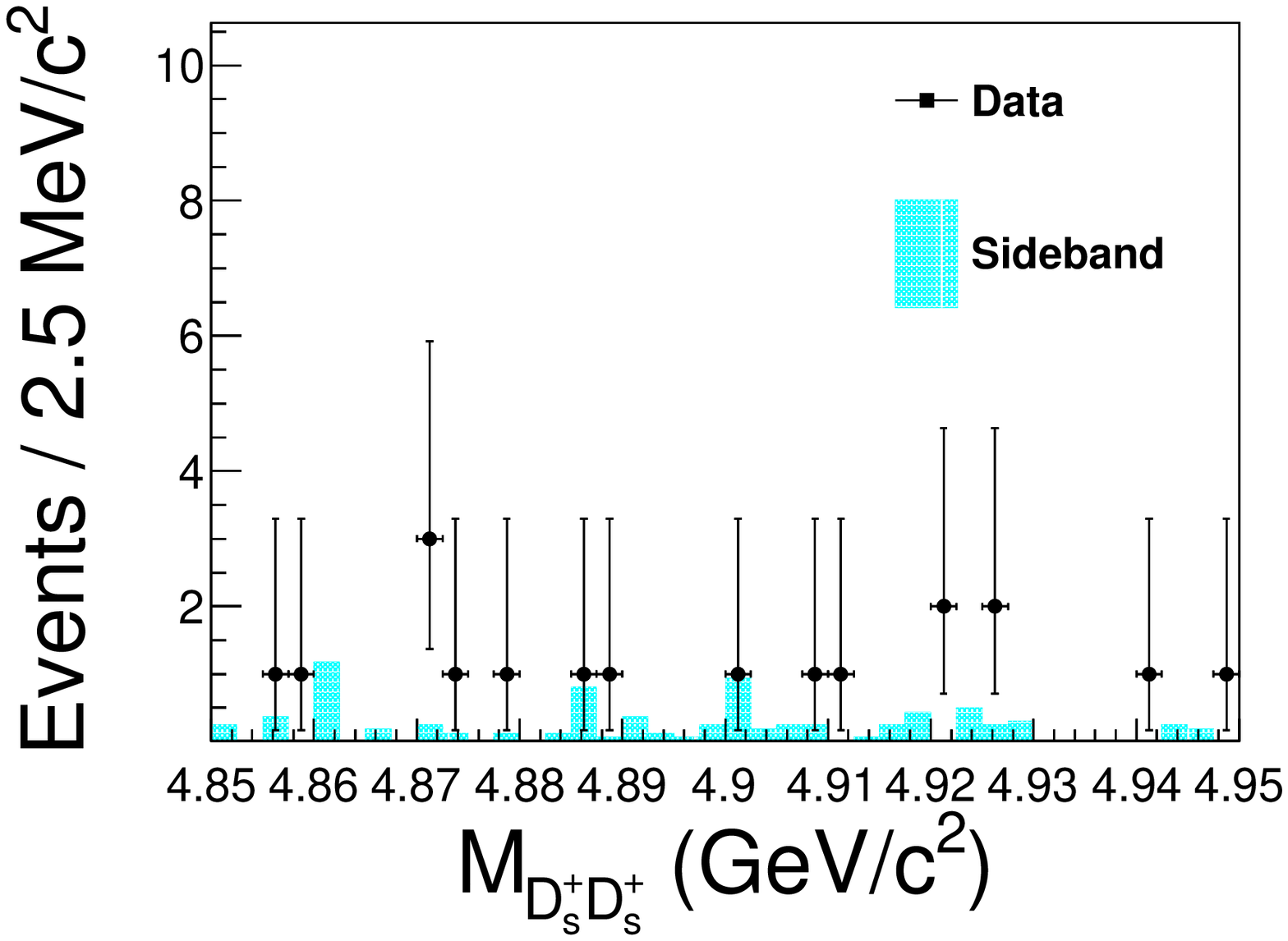}
    \put(-395,90){\bf (c)} \put(-250,90){\bf (d)} \put(-105,90){\bf (e)}
    \caption{Distributions of $M_{D_{s}^{+}D_{s}^{+}}$ from data for processes (a) $\ones \to \ccss (\to D_{s}^{+}D_{s}^{+}) + anything$, (b) $\twos \to \ccss (\to D_{s}^{+}D_{s}^{+}) + anything$, and $\EE \to \ccss (\to D_{s}^{+}D_{s}^{+}) + anything$ at (c) $\sqrt{s}$ = 10.52~GeV, (d) $\sqrt{s}$ = 10.58~GeV, (e) $\sqrt{s}$ = 10.867~GeV. The cyan shaded histograms are from the normalized $M_{D_{s}^{+}D_{s}^{+}}$ sideband events.}\label{DD_mass_data}
  \end{center}
\end{figure*}

\begin{figure*}[htbp]
  \begin{center}
    \includegraphics[height=3.75cm,width=5cm]{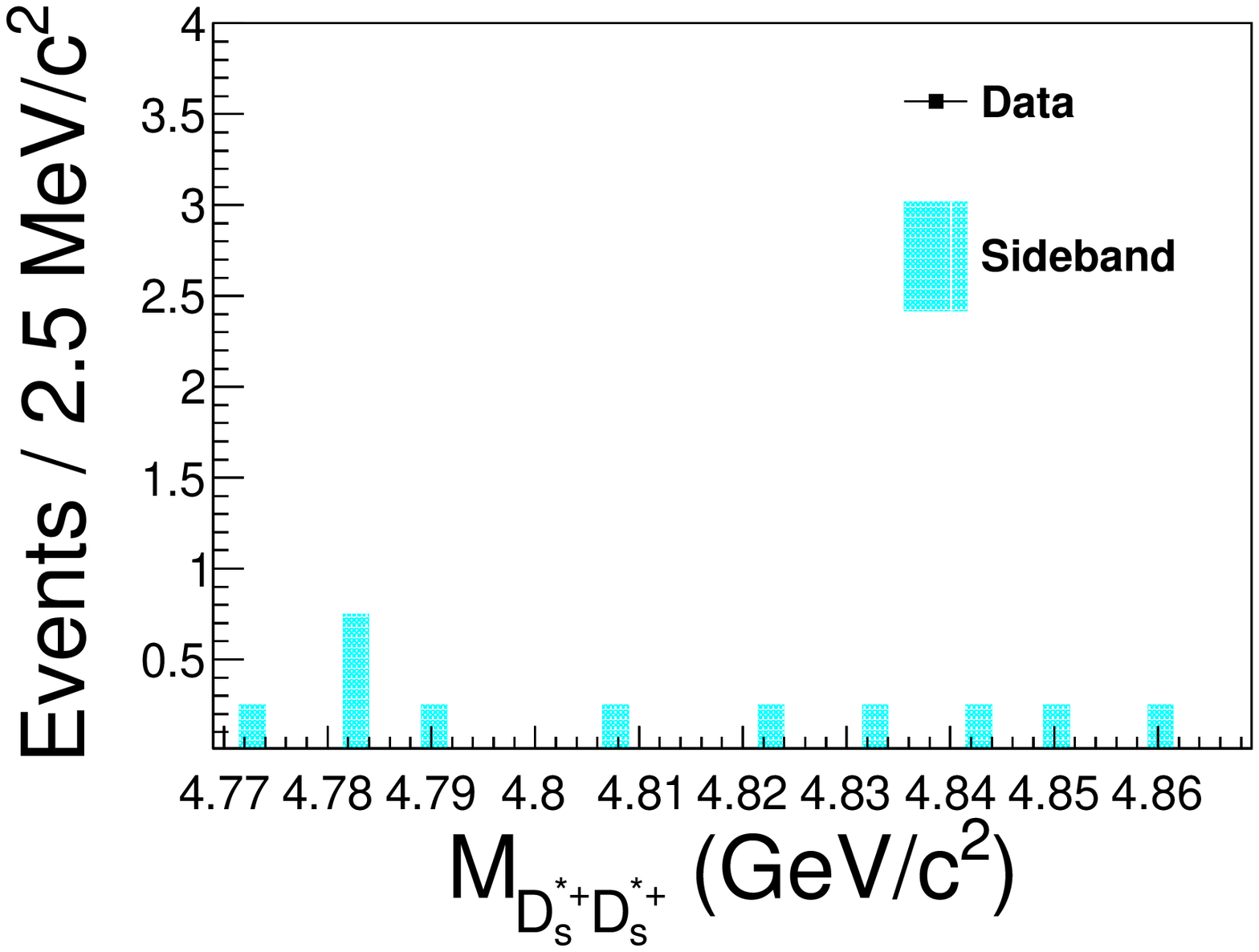}
    \includegraphics[height=3.75cm,width=5cm]{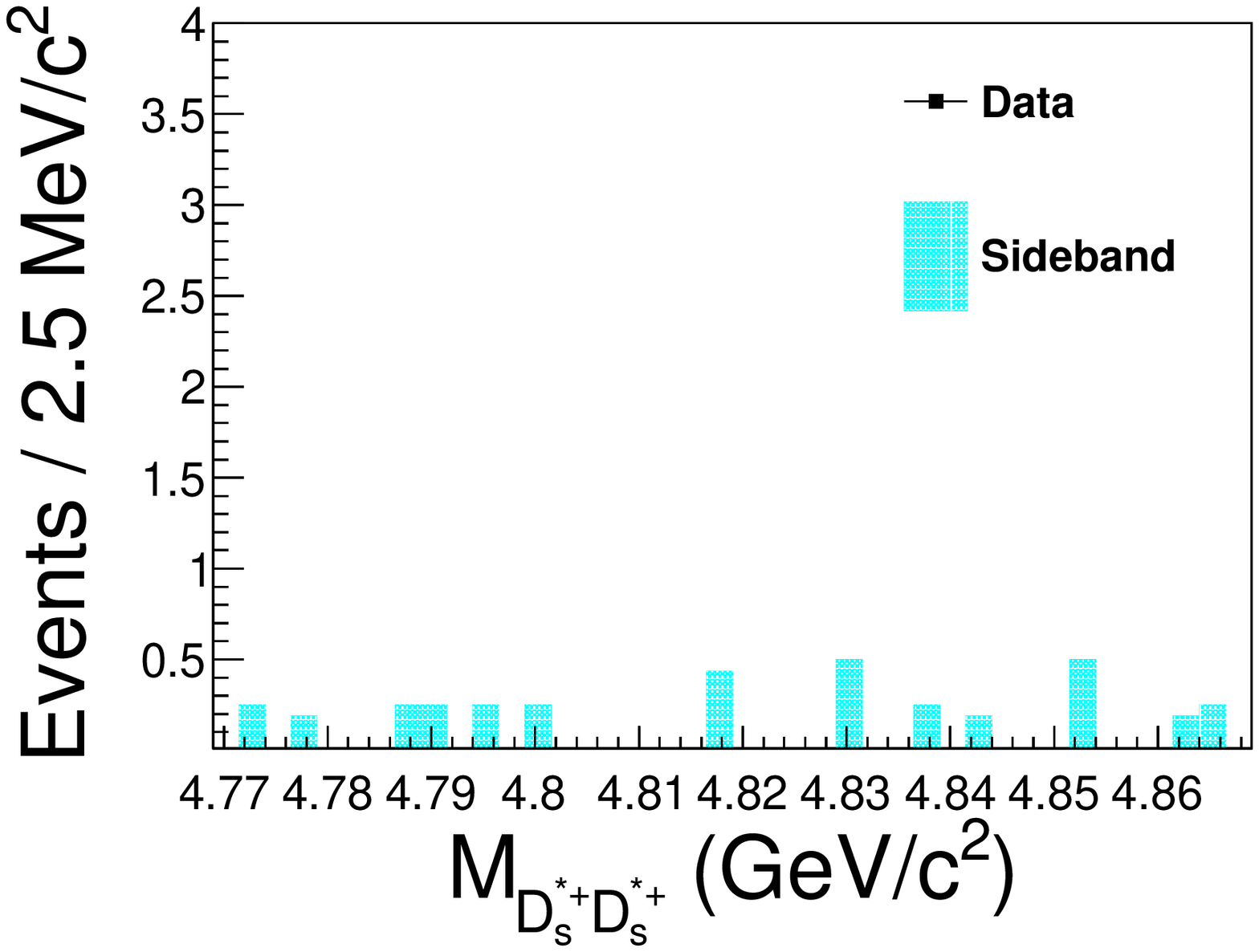}
    \put(-250,90){\bf (a)} \put(-105,90){\bf (b)}
\\
    \includegraphics[height=3.75cm,width=5cm]{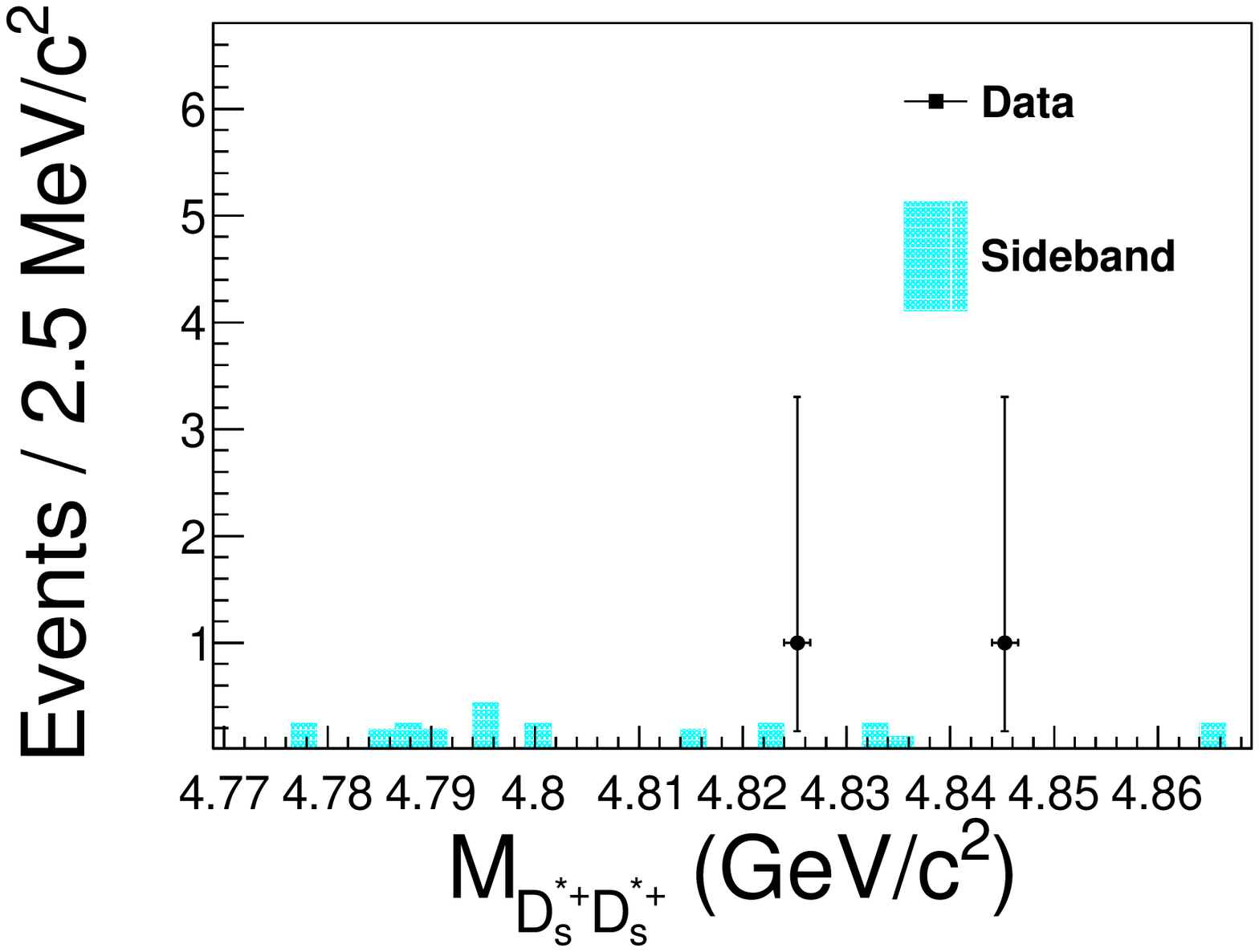}
    \includegraphics[height=3.75cm,width=5cm]{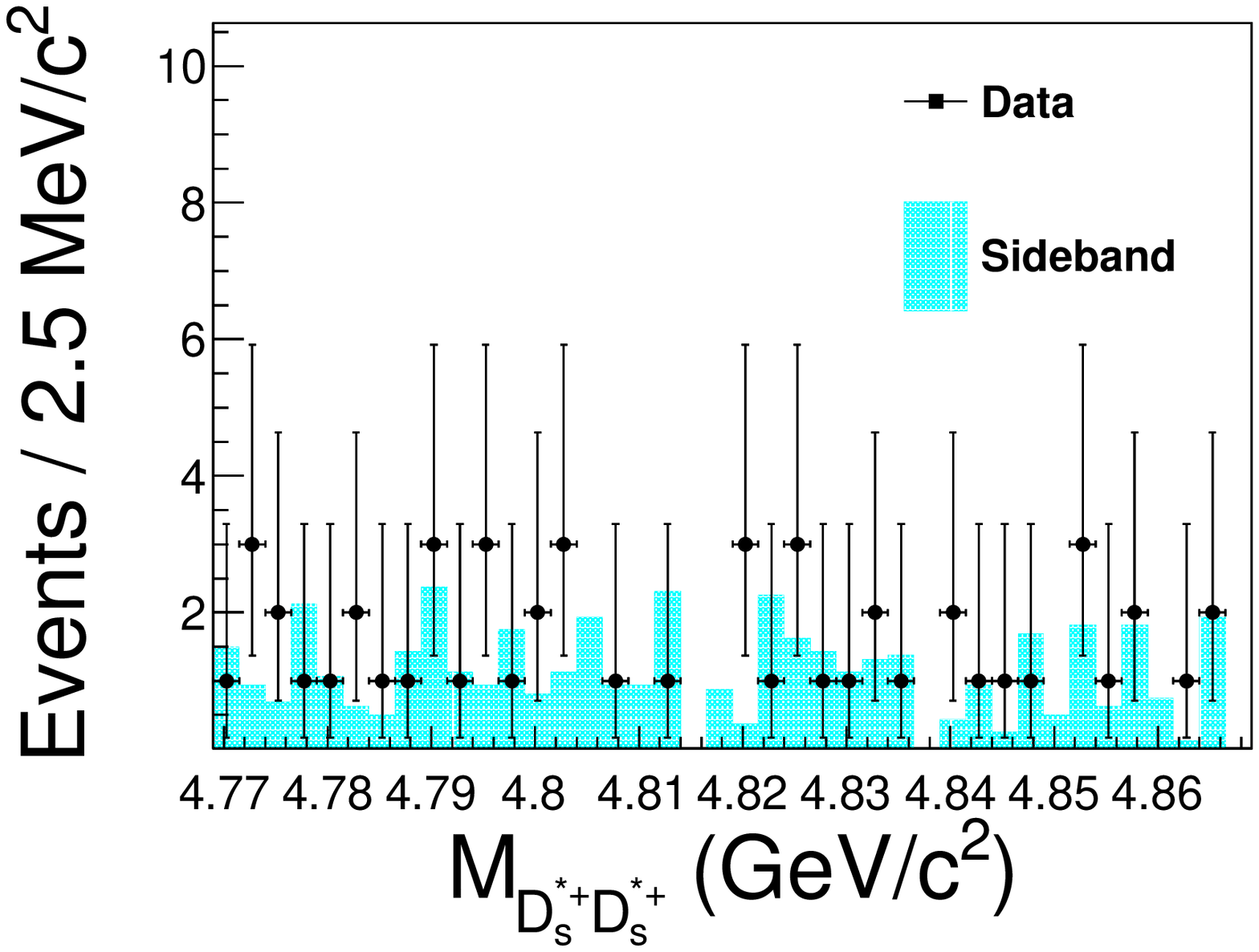}
    \includegraphics[height=3.75cm,width=5cm]{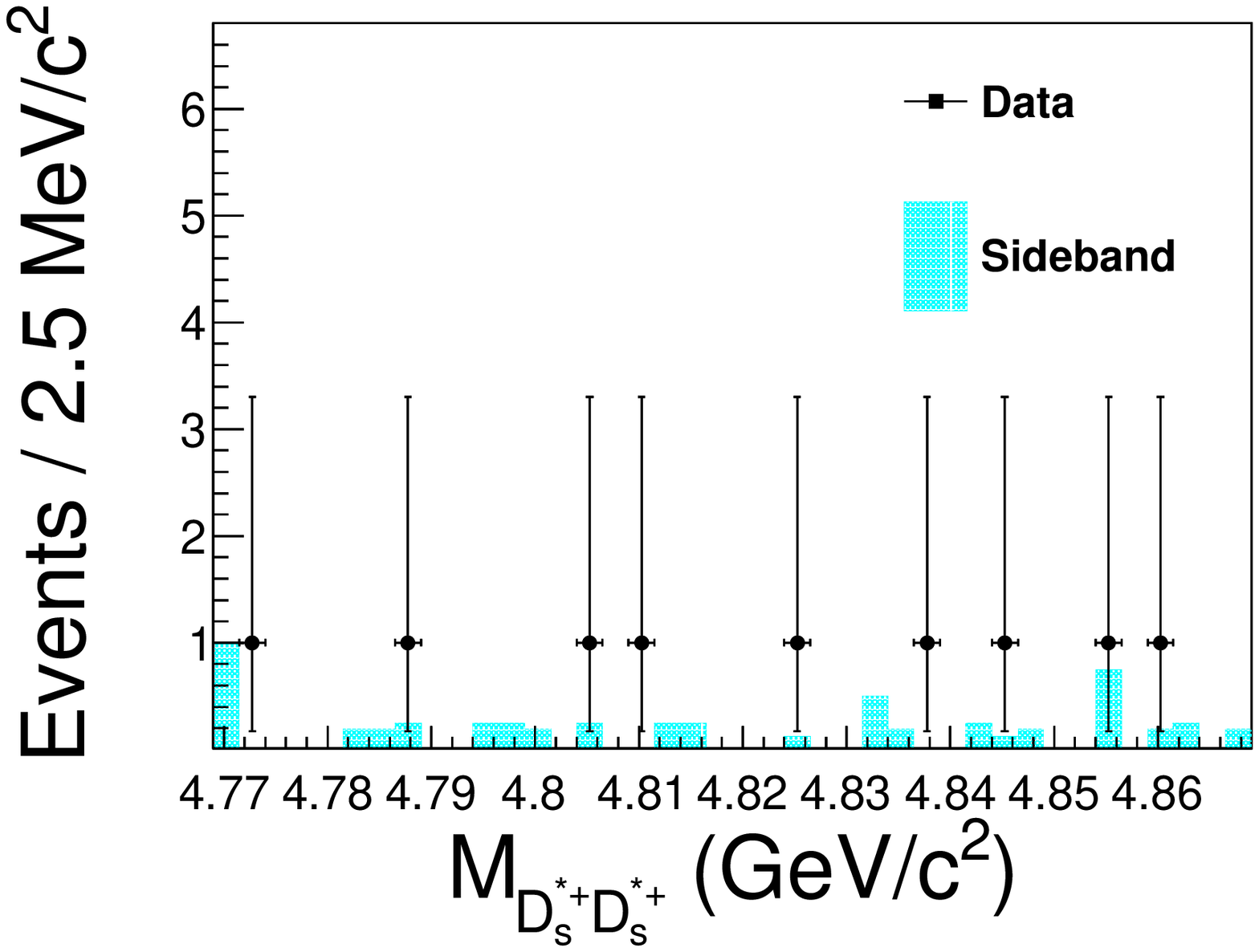}
    \put(-395,90){\bf (c)} \put(-250,90){\bf (d)} \put(-105,90){\bf (e)}
    \caption{Distributions of $M_{D_{s}^{*+}D_{s}^{*+}}$ from data for processes (a) $\ones \to \ccss (\to D_{s}^{*+}D_{s}^{*+}) + anything$, (b) $\twos \to \ccss (\to D_{s}^{*+}D_{s}^{*+}) + anything$, and $\EE \to \ccss (\to D_{s}^{*+}D_{s}^{*+}) + anything$ at (c) $\sqrt{s}$ = 10.52~GeV, (d) $\sqrt{s}$ = 10.58~GeV, (e) $\sqrt{s}$ = 10.867~GeV. The cyan shaded histograms are from the normalized $M_{D_{s}^{+}D_{s}^{+}}$ sideband events.}\label{DsDs_mass_data}
  \end{center}
\end{figure*}

The $D_{s}^{+}D_{s}^{+}$ and $D_{s}^{*+} D_{s}^{*+}$ invariant mass distributions of selected events from data samples in the kinematically allowed region are shown in Figs.~\ref{DD_mass_data_all} and ~\ref{DsDs_mass_data_all}
together with the backgrounds estimated from the normalized $D_{s}^{+}D_{s}^{+}$ sideband events.
No peaking backgrounds are found in the normalized sideband events in either $D_{s}^{+}D_{s}^{+}$ and $D_{s}^{*+}D_{s}^{*+}$ invariant mass distributions from data,
nor in the $D_{s}^{+}D_{s}^{+}$ and $D_{s}^{*+}D_{s}^{*+}$ mass spectra from inclusive MC samples~\cite{topoo}. Thus in the following we only focus on
the mass spectra from the theoretically predicted regions for $X_{cc\bar{s}\bar{s}}$~\cite{the_qqss2}
which are shown in Figs.~\ref{DD_mass_data} and ~\ref{DsDs_mass_data}.

Since no clear signals are observed in the invariant-mass spectra, the 90\% confidence level (C.L.) upper limits on the numbers of signal events
are given.  The upper limit is calculated by the frequentist approach~\cite{pole-1} implemented in the POLE (Poissonian limit estimator) program~\cite{pole-2}, where the mass window is obtained by giving 95\% acceptance to the corresponding simulated signal events, the number of signal candidate events is counted directly, and the number of expected background events is estimated from the normalized mass sidebands.
The possible non-resonant contributions in the $D_{s}^{+}D_{s}^{+}$ and $D_{s}^{*+}D_{s}^{*+}$ invariant-mass spectra are not subtracted and
taken as potential signals, in order to set more conservative upper limits.



The upper limit calculation is repeated with $M_{\ccss}$ varying from 4882 MeV/$c^2$ to 4922 MeV/$c^2$ in steps of 5~MeV/$c^2$ and $\Gamma_{\ccss}$ varying
from 0.54~MeV to 6.54~MeV in steps of 1.0~MeV for the $M_{D_{s}^{+}D_{s}^{+}}$ distribution, and with $M_{\ccss}$ varying from 4801 MeV/$c^2$ to 4841 MeV/$c^2$ in steps of 5 MeV/$c^2$
and $\Gamma_{\ccss}$ varying from 2.58~MeV to 8.58~MeV in steps of 1.0 MeV for the $M_{D_{s}^{*+}D_{s}^{*+}}$ distribution.

\section{\boldmath Systematic Uncertainties }
There are several sources of systematic uncertainties on the branching fraction and Born cross section measurements,
which can be divided into multiplicative and additive systematic uncertainties.
The multiplicative systematic uncertainties include detection-efficiency-related (DER) sources (tracking efficiency,
PID, and photon reconstruction),
the statistical uncertainty of the MC efficiency, branching fractions of intermediate states, the total numbers of $\ones$ and $\twos$
events, and the integrated luminosities at $\sqrt{s}$ = 10.52~GeV, 10.58~GeV, and 10.867~GeV.

The systematic uncertainties related to detection efficiency ($\sigma_{\textrm{DER}}$) include the tracking efficiency (0.35\% per track, estimated using partially reconstructed $D^{\ast}$ decays in $D^{*+} \to \pi^{+} D^0, D^0 \to K_S^0 \pi^{+} \pi^{-}$), PID efficiency ($2.2\%$ per kaon and $1.8\%$ per pion, estimated using $D^{*+} \to D^{0}\pi^{+}$, $D^{0} \to K^{-} \pi^{+}$ samples), and photon reconstruction (2.0\% per photon, estimated using a radiative Bhabha sample).
The statistical uncertainty in the signal MC simulation efficiency can be calculated as $\Delta \varepsilon$ = $\sqrt{\varepsilon(1-\varepsilon)/N}$, where $\varepsilon$ is the reconstruction efficiency after all event selections, and $N$ is the total number of generated events.
Its relative uncertainty $\sigma_{\textrm{MC stat.}} = \Delta\varepsilon/\varepsilon$ is at most at the 1.0\% level.
Changing the $s$ dependence of the cross sections of $\EE \to \ccss (\to D_{s}^{*+}D_{s}^{*+}) + anything$ from $1/s$ to $1/s^{4}$, the product of efficiency and radiative correction factor $\epsilon(1+\delta)_{\textrm{ISR}}$ changes by less than 0.3\% ($\sigma_{\textrm{ISR}}$).

The relative uncertainties of branching fractions for $D_{s}^{*+} \to \gamma D_{s}^{+}$, $D_{s}^+ \to \phi(\to K^{+}K^{-})\pip$, and $D_{s}^+ \to \bar{K}^{*}(892)^{0}(\to K^{-}\pip)K^{+}$ are 0.75\%, 3.52\%, and 3.45\%~\cite{PDG}, respectively.
The total uncertainties are calculated using  $\sigma_{\BR} = \frac{\sqrt{\Sigma{(\varepsilon_{i} \times \BR_i \times \sigma_{\BR_{i}})^2}}}{\Sigma{(\varepsilon_{i} \times \BR_{i}})}$, where $\varepsilon_{i}$ is the efficiency, $\sigma_{\BR_i}$ is the relative uncertainty of intermediate states' branching fractions, and $\BR_i$
is the product of branching fractions of the intermediate states for each reconstructed mode $i$.

The total numbers of $\ones$ and $\twos$ events are estimated to be ($102 \pm 2) \times 10^6$ and ($157.8 \pm 3.6) \times 10^6$, which are determined by counting the numbers of inclusive hadrons.
The uncertainties are mainly due to imperfect simulations of the charged multiplicity distributions from inclusive hadronic MC events ($\sigma_{\textrm{N}_{\onetwos}}$).
Belle measures luminosity with 1.4\% precision using wide angle Bhabha events ($\sigma_{\lum}$).

All the multiplicative uncertainties are summarized in Table~\ref{tab:DsDs_err} for the measurements of $\onetwos \to \ccss + anything$ and $\EE \to \ccss + anything$ at $\sqrt{s} $ = 10.52~GeV, 10.58~GeV, and 10.867 GeV, respectively. The total multiplicative uncertainty is calculated by adding all sources of multiplicative uncertainty in quadrature,$$\sigma_{\textrm{syst.}} = \sqrt{\sigma_{\textrm{DER}}^2 + \sigma_{\textrm{MC stat.}}^2 + \sigma_{\textrm{ISR}}^2 + \sigma_{\BR}^2 + \sigma_{\textrm{N}_{\onetwos / \lum}}^2}.$$
The additive uncertainty due to the number of expected background is considered by counting normalized background distributions directly, fitting the distributions with a constant, and a 1st-order polynominal.

\begin{table*}[htbp]
  \caption{\label{tab:DsDs_err}  Summary of the multiplicative systematic uncertainties (\%) on the branching fraction measurements for $\onetwos \to \ccss(\to D_{s}^{+}D_{s}^{+}(D_{s}^{*+}D_{s}^{*+})) + anything$ and on the Born cross section measurements for $\EE \to \ccss(\to D_{s}^{+}D_{s}^{+}(D_{s}^{*+}D_{s}^{*+})) + anything$ at $\sqrt{s} $ = 10.52~GeV, 10.58~GeV, and 10.867~GeV.}
  \small
  \begin{tabular}{ccccccc}
      \hline
      \hline
     \tabincell{c}{$M_{D_{s}^{+}D_{s}^{+}}$ ($M_{D_{s}^{*+}D_{s}^{*+}}$) mode} &  DER & MC stat.  & \tabincell{c}{\textrm{ISR}}  & $\BR$ & \tabincell{c}{ [$\textrm{N}_{\ones}/\textrm{N}_{\twos}/\lum $]}& \tabincell{c}{\textrm{Sum}} \\
      \hline
     $\ones \to \ccss + anything$                          & 6.1 (7.3) & 1.0 &---  & 3.0 & 2.0 & 7.2 (8.2)    \\
       $\twos \to \ccss + anything$                          & 6.1 (7.3) & 1.0 &---  & 3.0 & 2.3 & 7.2 (8.3)    \\
       $\EE \to \ccss + anything$ at $\sqrt{s}$ = 10.52 GeV  & 6.1 (7.3) & 1.0 &0.3  & 3.0 & 1.4 & 7.0 (8.2)    \\
       $\EE \to \ccss + anything$ at $\sqrt{s}$ = 10.58 GeV  & 6.1 (7.3) & 1.0 &0.3  & 3.0 & 1.4 & 7.0 (8.2)    \\
       $\EE \to \ccss + anything$ at $\sqrt{s}$ = 10.867 GeV & 6.1 (7.3) & 1.0 &0.3  & 3.0 & 1.4 & 7.0 (8.2)    \\
      \hline
      \hline
  \end{tabular}
\end{table*}

\section{Statistical interpretation of upper limit setting}
Since no signal traces are observed in the $D_{s}^{+}D_{s}^{+}$ or $D_{s}^{*+} D_{s}^{*+}$ distributions from data at all energy points,
the 90\% C.L.\ upper limits on the numbers of signal events ($N^{\textrm{UP}}$) are determined.
To take into account the additive and multiplicative uncertainties, we first study the additive systematic uncertainty and take the most conservative case, then use the total multiplicative systematic uncertainty as an input parameter to the POLE program.

Since there are few events observed from data sample at $\sqrt{s}$ = 10.52 GeV, the continuum contributions are neglected for the $\onetwos$ decays.
The conservative upper limit on the product branching fractions in $\onetwos$ decays $\BR^{\textrm{UP}}(\onetwos \to \ccss + anything) \times \BR(\ccss \to D_{s}^{+}D_{s}^{+}(D_{s}^{*+} D_{s}^{*+}))$ are obtained by the following formula:
$$\frac{N^{\textrm{UP}}}{N_{\onetwos} \times \sum_{i}\varepsilon_{i}\BR_{i} },$$
where $N^{\textrm{UP}}$ is the 90\% C.L.\ upper limit on the number of events from the data signal yields including all systematic uncertainties that are mentioned above from other variables in this expression, $N_{\onetwos}$ is the total number of $\onetwos$ events,
$\varepsilon_{i}$ is the corresponding detection efficiency, and $\BR_{i}$ is the product of all secondary branching fractions for each reconstructed channel.

The conservative upper limit on the product values of Born cross section and branching fraction $\sigma^{\textrm{UP}}(\EE \to \ccss + anything) \times \BR(\ccss \to D_{s}^{+}D_{s}^{+}(D_{s}^{*+} D_{s}^{*+}))$ are calculated by the following formula:
$$\frac{N^{\textrm{UP}} \times |1-\Pi|^{2}}{\lum \times \sum_{i}\varepsilon_{i}\BR_{i}
\times (1+\delta)_{\textrm{ISR}}},$$
where $N^{\textrm{UP}}$ is the 90\% C.L.\ upper limit on the number of events in data signal yields including all systematic uncertainties that are mentioned above from other variables in this expression, $|1-\Pi|^{2}$ is the vacuum polarization factor, $\lum$ is the integrated luminosity, $\varepsilon_{i}$ is the corresponding detection efficiency, $\BR_{\textrm{i}}$ is the product of all secondary branching fractions for each reconstructed channel, and $(1+\delta)_{\textrm{ISR}}$ is the radiative correction factor.
The values of $|1-\Pi|^{2}$ are 0.931, 0.930, and 0.929 for $\sqrt{s}$ = 10.52~GeV, 10.58~GeV, and 10.867~GeV~\cite{vacuum}, and the uncertainty is calculated to be less than 0.1\%, which is negligible.
The radiative correction factors $(1+\delta)_{\textrm{ISR}}$ are 0.686, 0.694, and 0.738, as calculated using the formula given in Ref.~\cite{ISR} for $\sqrt{s}$ = 10.52~GeV, 10.58~GeV, and 10.867~GeV, respectively, where we assume that the dependence of cross sections on $s$ is $1/s$.

\begin{figure*}[!htbp]
  \begin{center}
    \includegraphics[height=5.4cm,width=5.5cm]{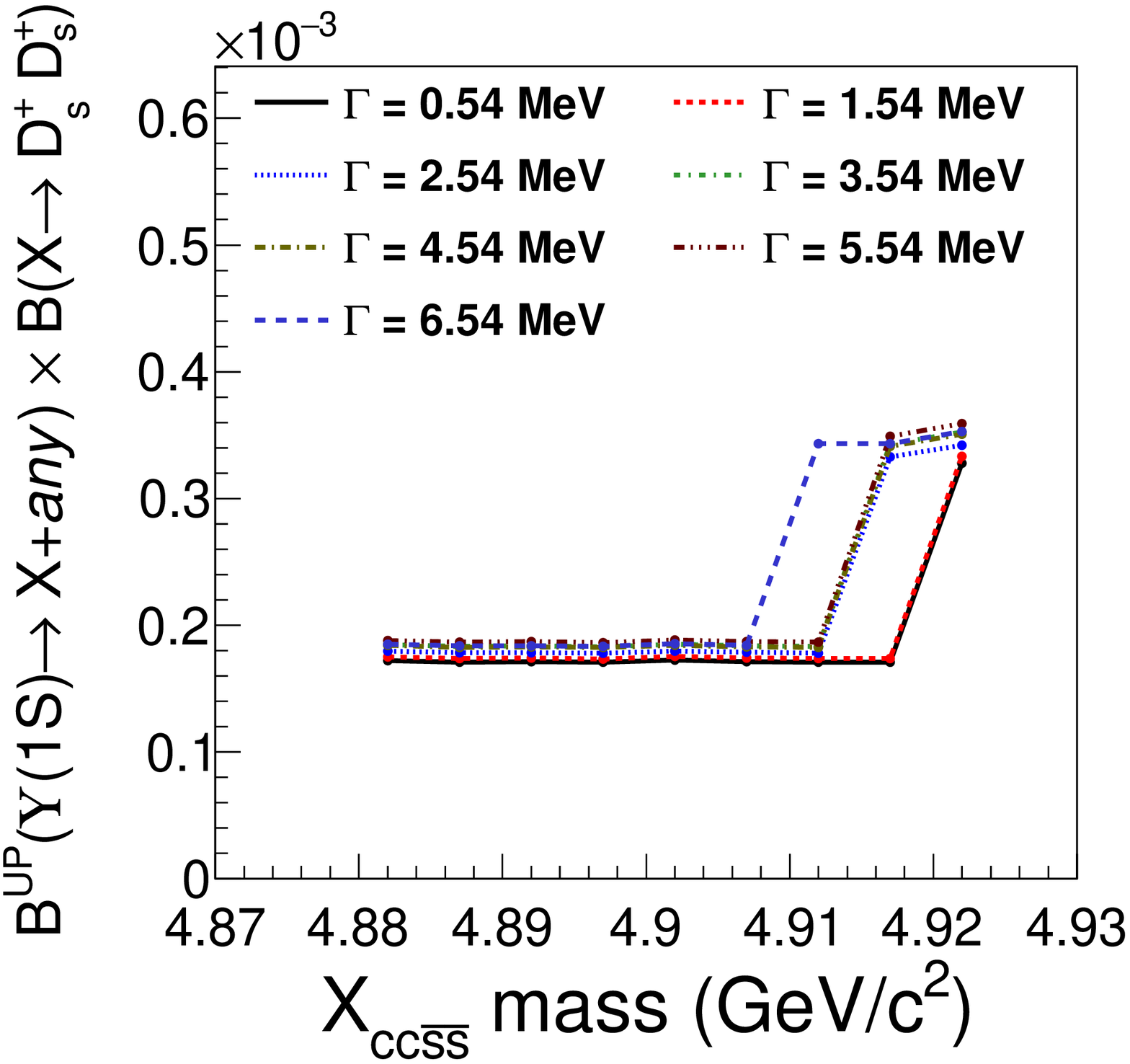}
    \includegraphics[height=5.4cm,width=5.5cm]{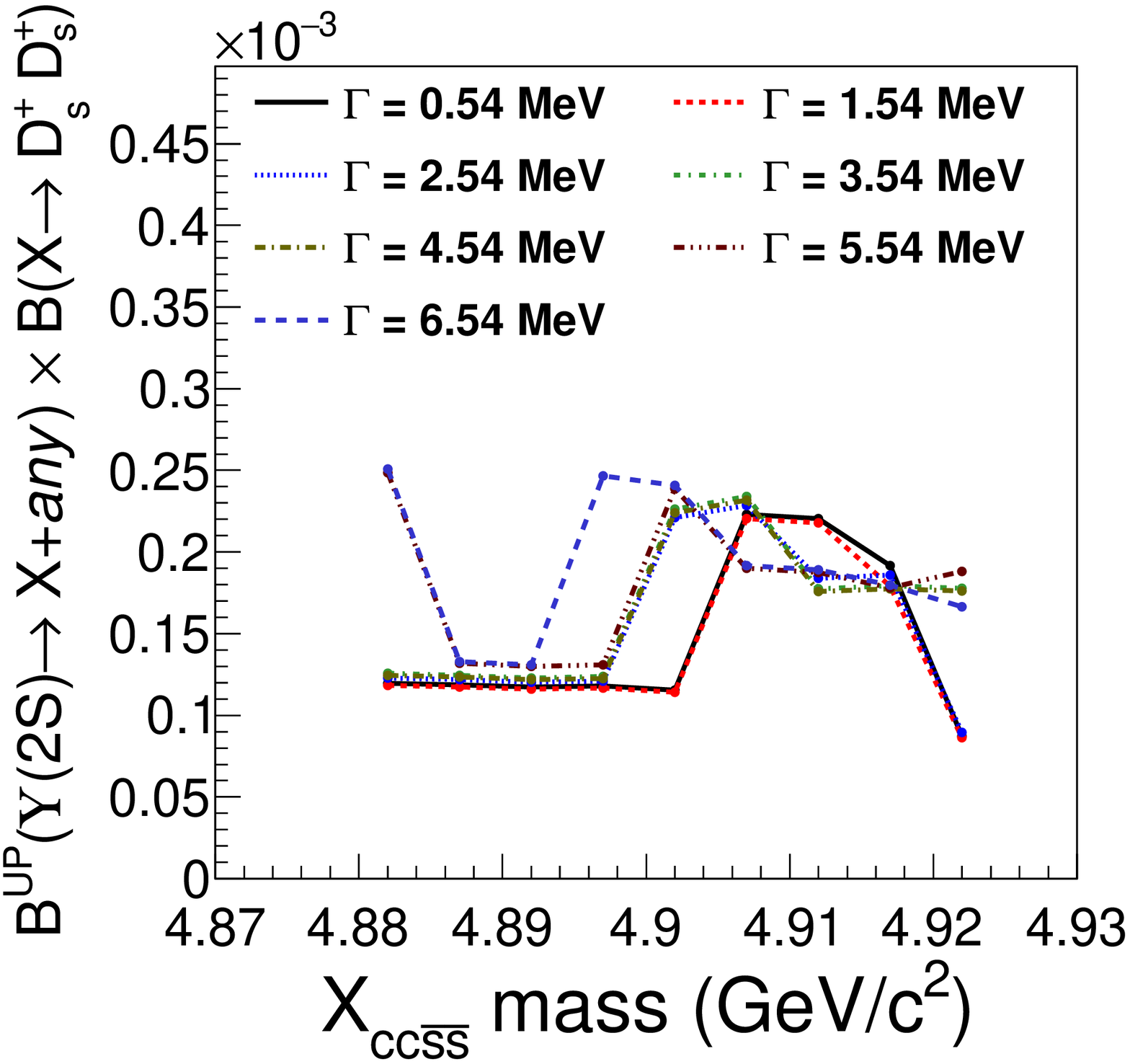}
\\
    \includegraphics[height=5.4cm,width=5.5cm]{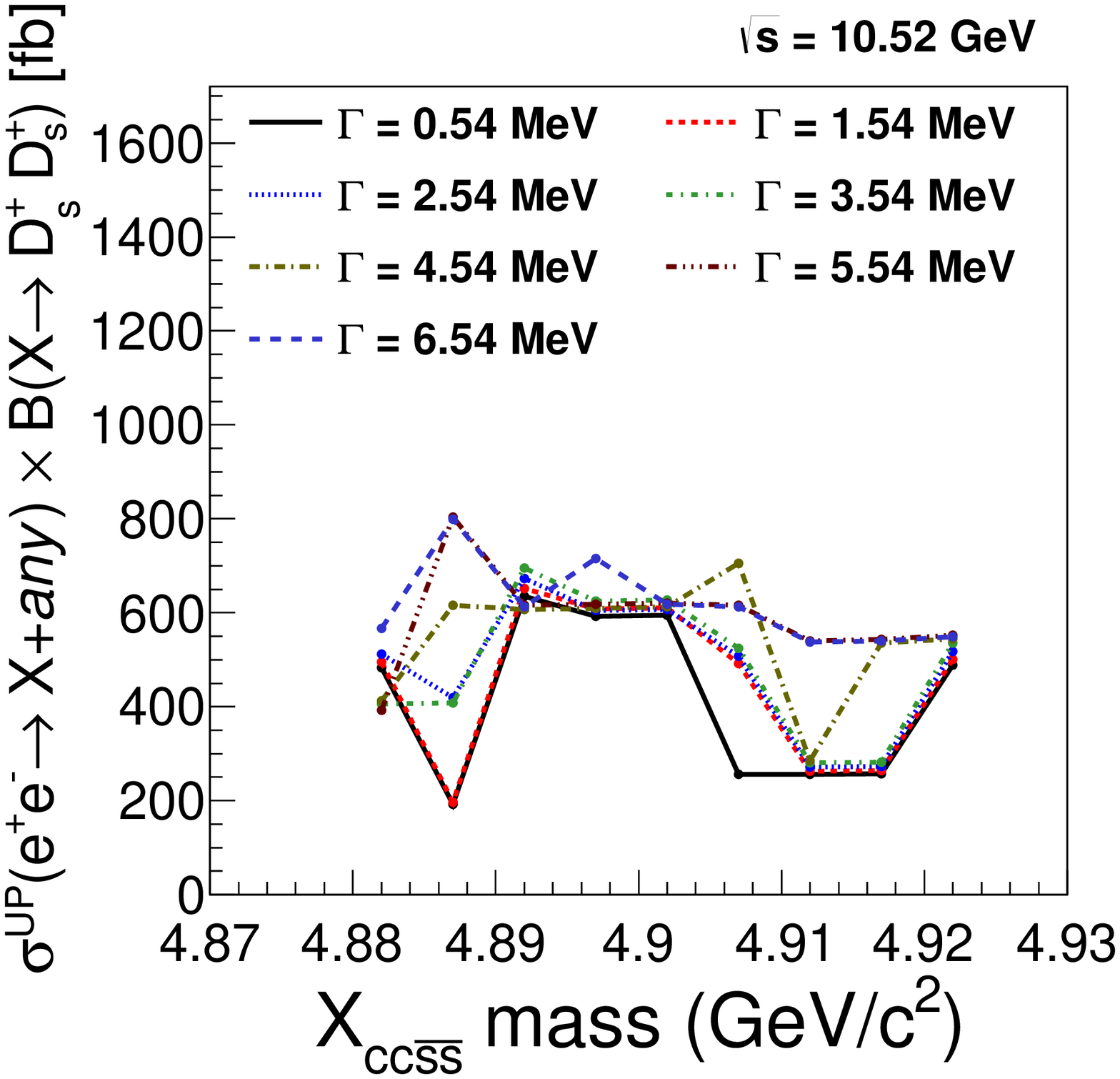}
    \includegraphics[height=5.4cm,width=5.5cm]{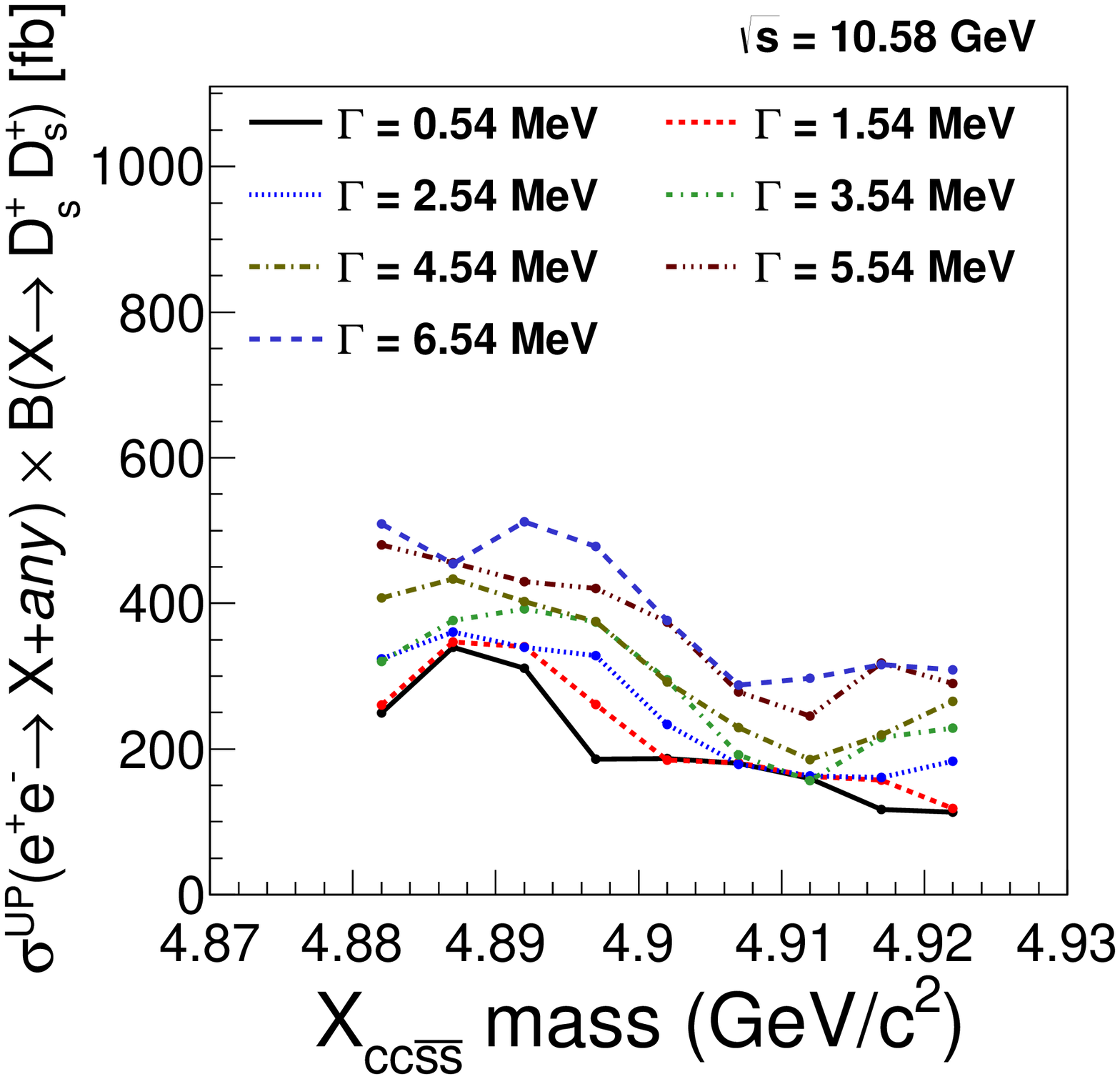}
    \includegraphics[height=5.4cm,width=5.5cm]{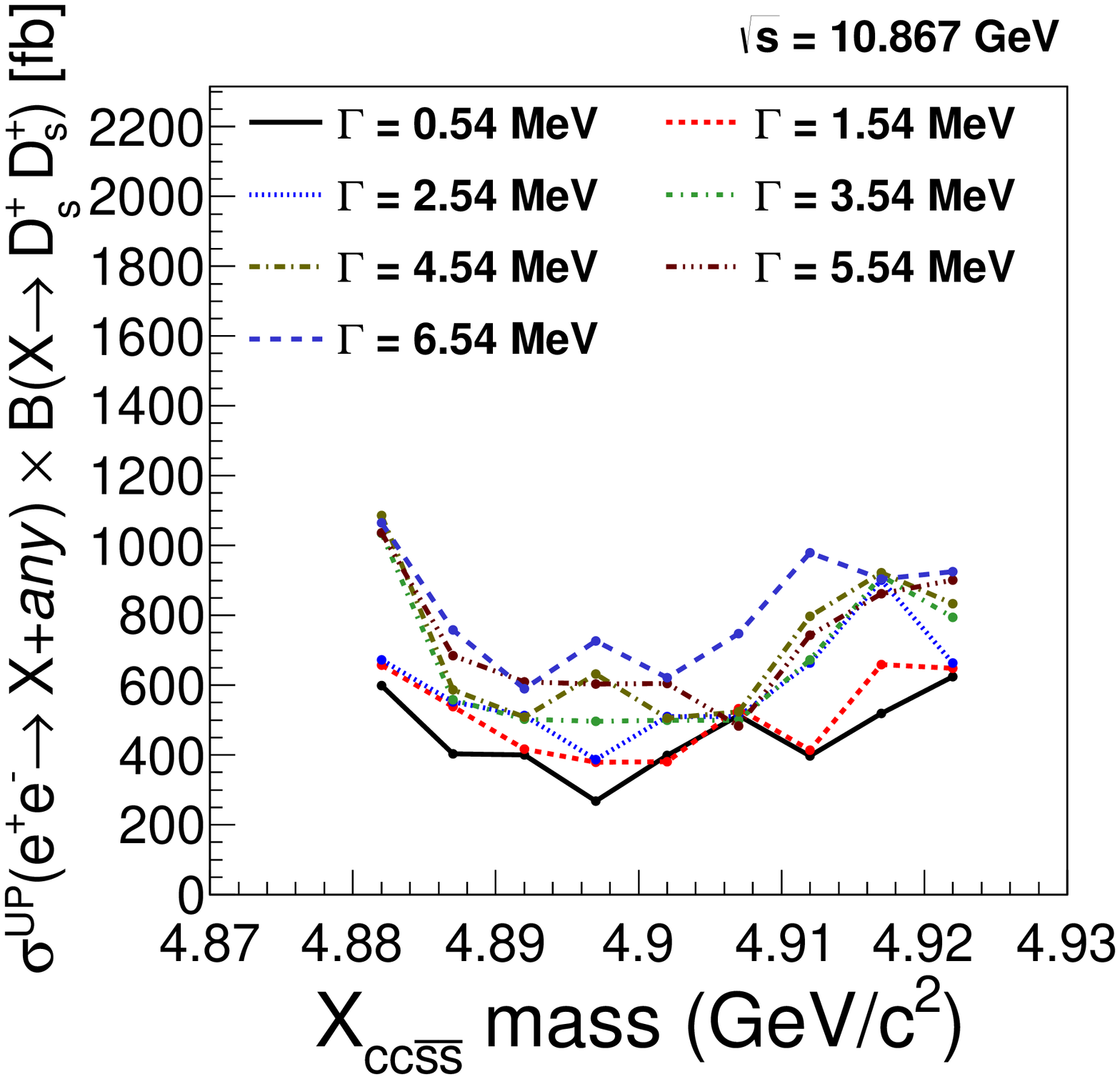}
    \caption{The 90\% C.L. upper limits on the product branching fractions of $\onetwos \to \ccss(\to \dsds) + anything$ and the Born cross sections of $\EE \to \ccss + anything$ at $\sqrt{s}$ = 10.52~GeV, 10.58~GeV, and 10.867~GeV with $M_{\ccss}$ varying from 4882~MeV/$c^2$ to 4922~MeV/$c^2$ in steps of 5~MeV/$c^2$ and $\Gamma_{\ccss}$ varying from 0.54~MeV to 6.54~MeV in steps of 1.0~MeV.}\label{DD_Xup}
  \end{center}
\end{figure*}

\begin{figure*}[htbp]
  \begin{center}
    \includegraphics[height=5.4cm,width=5.5cm]{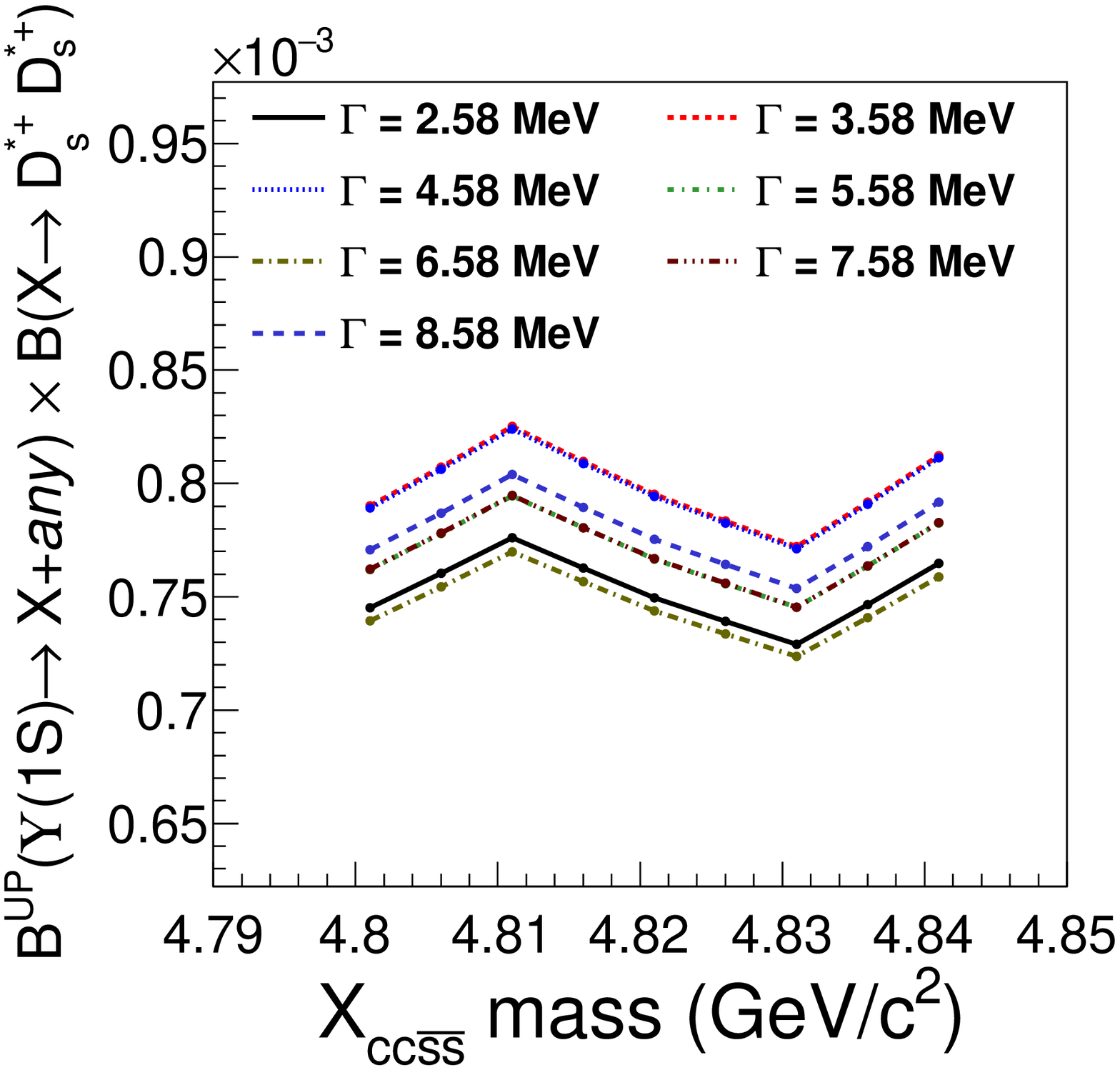}
    \includegraphics[height=5.4cm,width=5.5cm]{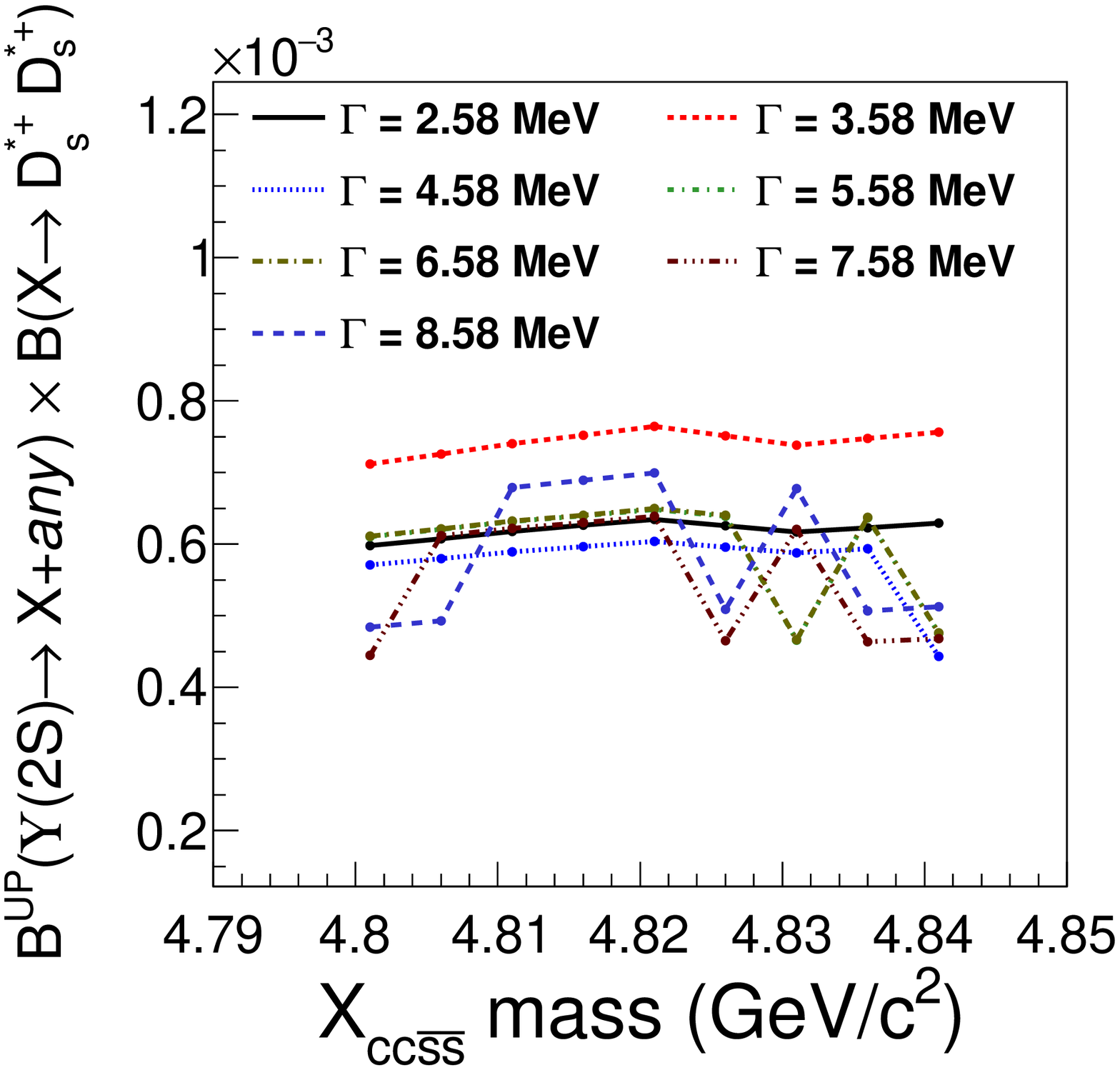}
\\
    \includegraphics[height=5.4cm,width=5.5cm]{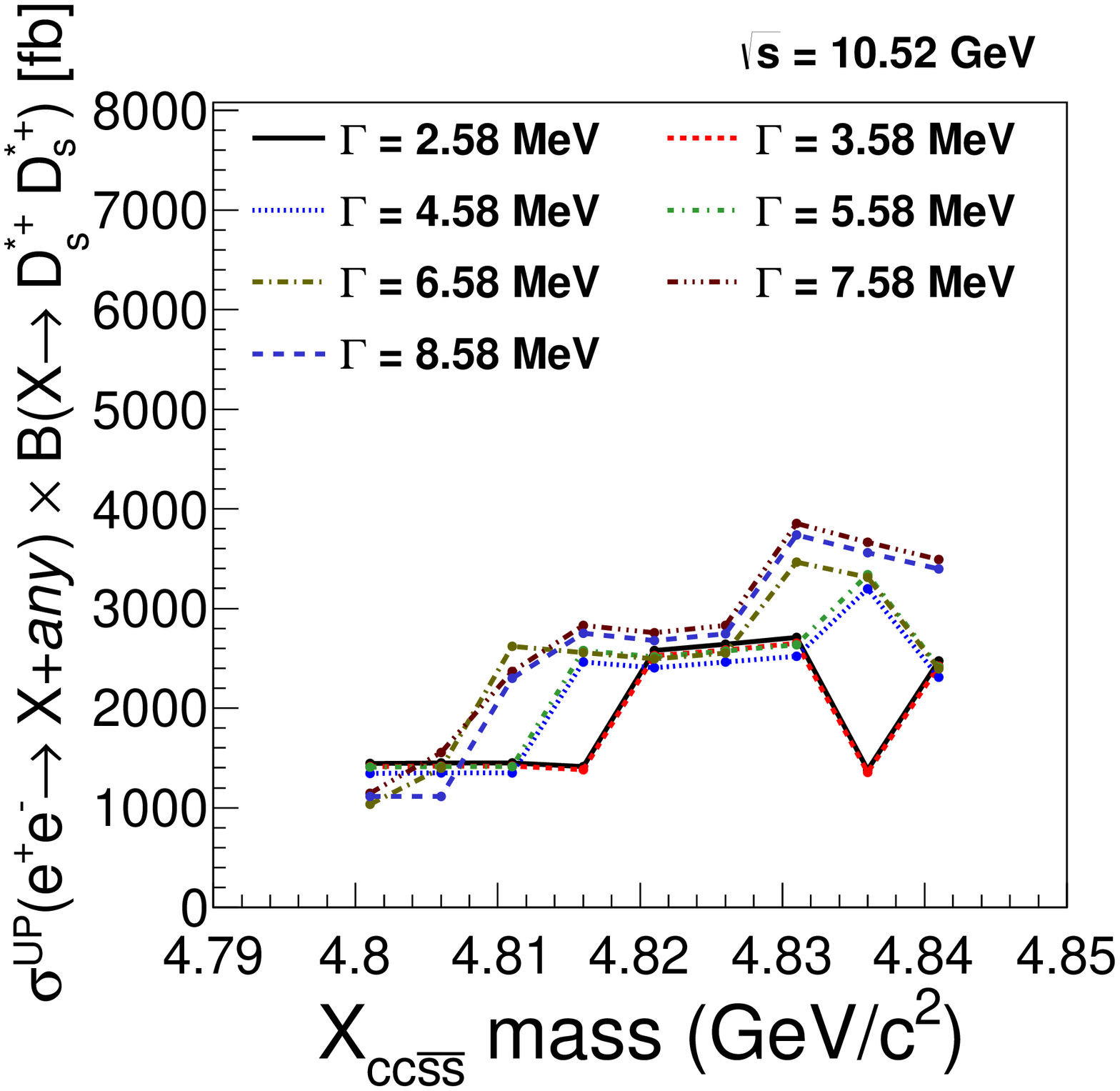}
    \includegraphics[height=5.4cm,width=5.5cm]{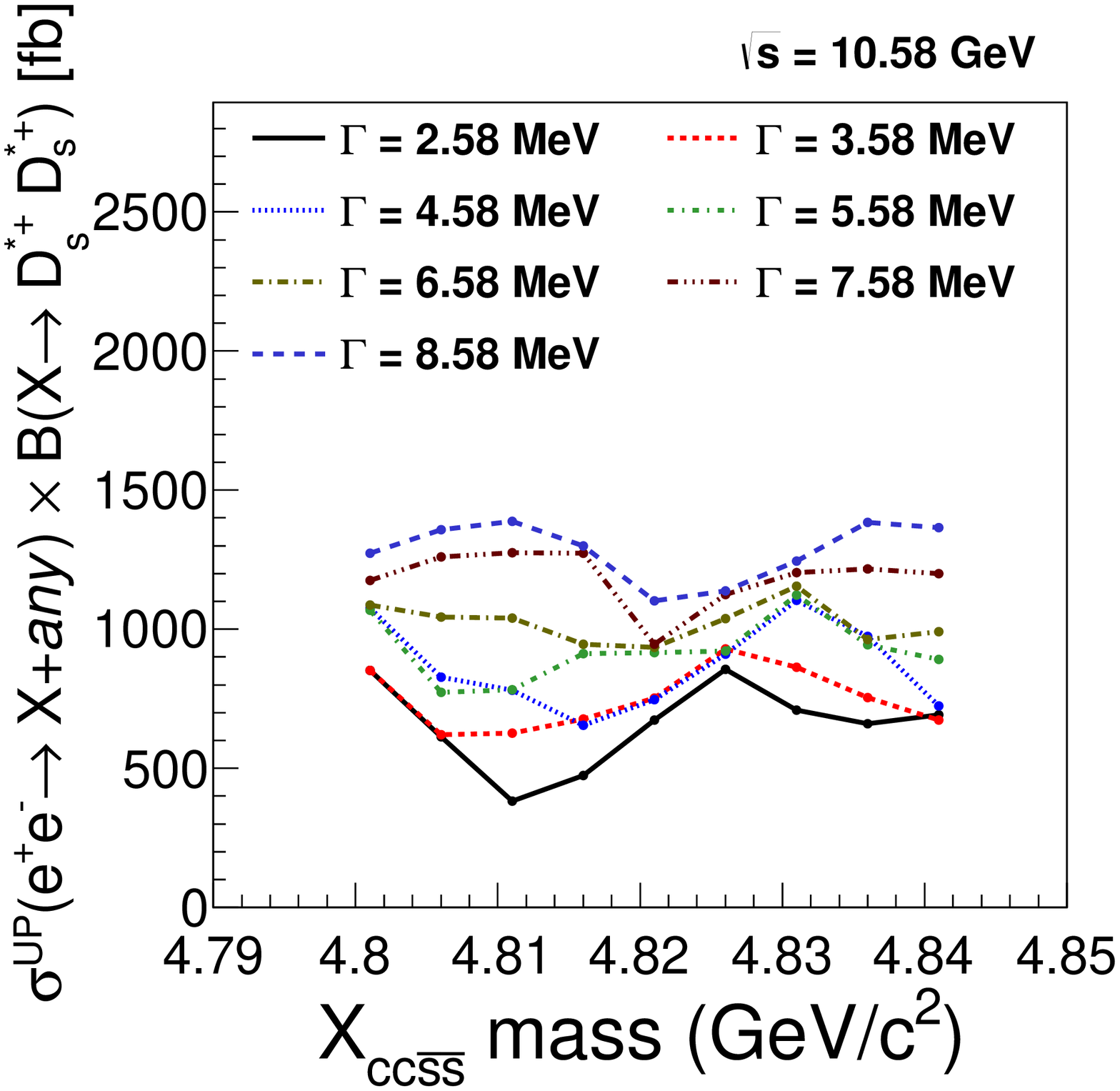}
    \includegraphics[height=5.4cm,width=5.5cm]{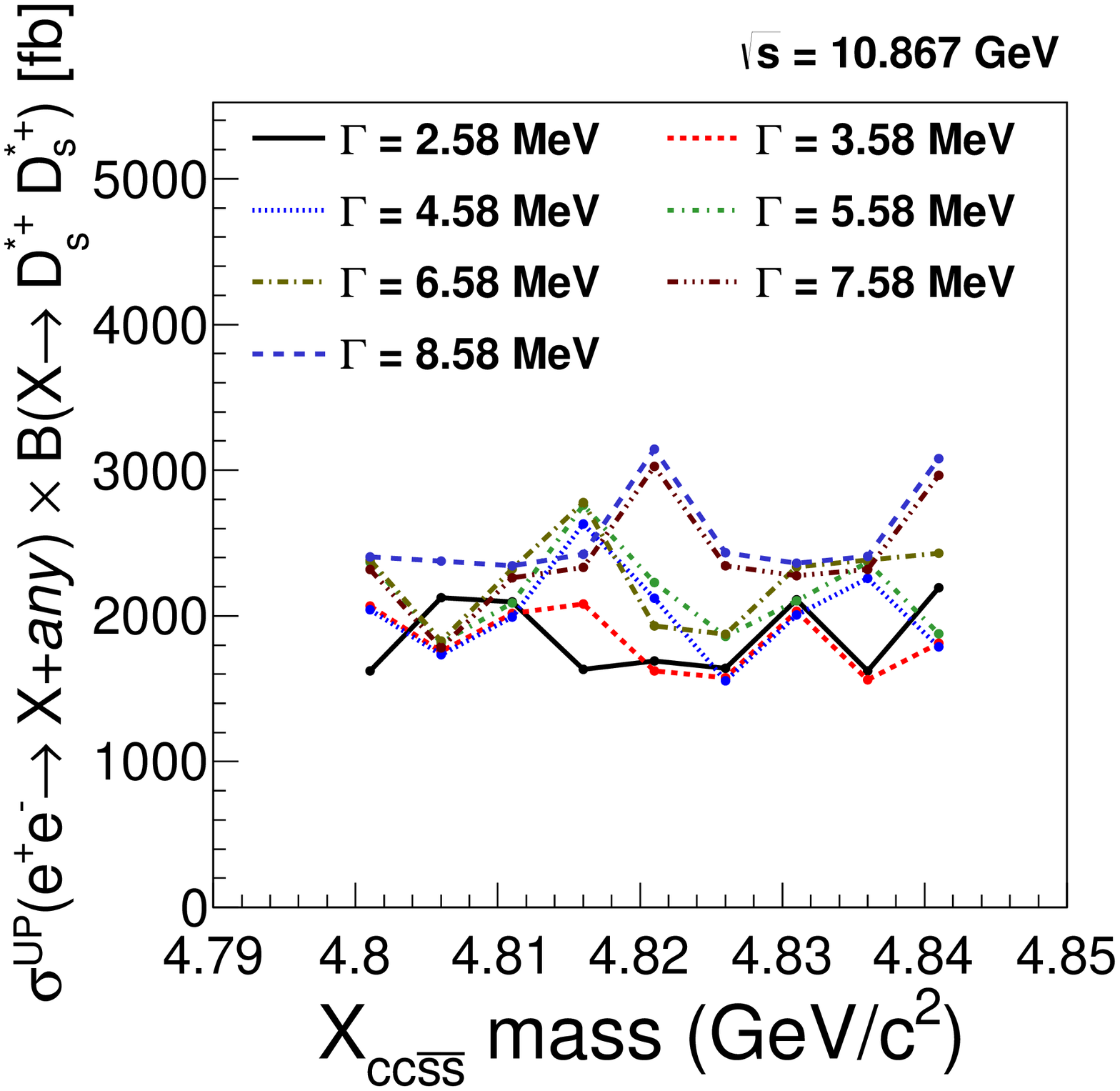}
    \caption{The 90\% C.L. upper limits on the product branching fractions of $\onetwos \to \ccss(\to \dstdst) + anything$ and the Born cross sections of $\EE \to \ccss + anything$ at $\sqrt{s}$ = 10.52~GeV, 10.58~GeV, and 10.867~GeV with $M_{\ccss}$ varying from 4801~MeV/$c^2$ to 4841~MeV/$c^2$ in steps of 5~MeV/$c^2$ and $\Gamma_{\ccss}$ varying from 2.58~MeV to 8.58~MeV in steps of 1.0~MeV.}\label{DsDs_Xup}
  \end{center}
\end{figure*}

The calculated 90\% C.L. upper limits on the product branching fractions of $\onetwos \to \ccss + anything$ and the product values of Born cross section and branching fraction of $\EE \to \ccss + anything$ at $\sqrt{s}$ = 10.52~GeV, 10.58~GeV, and 10.867~GeV for the mode $\ccss \to \dsds$ ($\ccss \to \dstdst$)
are displayed in Fig.~\ref{DD_Xup} (\ref{DsDs_Xup}).
Numerical values for the mode $\ccss \to \dsds$ can be found in Tables~\ref{tab:Ds_Ds_Xup_part1} and~\ref{tab:Ds_Ds_Xup_part2}, while those for the mode $\ccss \to \dstdst$ are shown in Tables~\ref{tab:Dstar Dstar_Xup_part1} and~\ref{tab:Dstar Dstar_Xup_part2}.

\begin{table*}[htpb]\scriptsize
  \begin{center}
    \caption{\label{tab:Ds_Ds_Xup_part1} Summary of 90\% C.L. upper limits with the systematic uncertainties included on the product branching fractions of $\ones/\twos \to \ccss (\to D_{s}^{+}D_{s}^{+}) + anything$. }
    \renewcommand\arraystretch{1.3}
\begin{tabular}{cccccccc}
      \hline
      \hline
\multicolumn{8}{c}{$\BR(\ones / \twos \to \ccss + anything) \times \BR(\ccss \to D_{s}^{+}D_{s}^{+})$ ($\times10^{-4}$)}                                                                                                                                                                                                                                   \\
      \hline
\multicolumn{1}{c}{\textbf{$M_{\ccss}$ (MeV/$c^{2}$)}} &\multicolumn{7}{c}{\textbf{$\Gamma_{\ccss}$ (MeV)}} \\

 & \multicolumn{1}{c}{\textbf{$0.54$}} & \multicolumn{1}{c}{\textbf{$1.54$}} & \multicolumn{1}{c}{\textbf{$2.54$}} & \multicolumn{1}{c}{\textbf{$3.54$}} & \multicolumn{1}{c}{\textbf{$4.54$}} & \multicolumn{1}{c}{\textbf{$5.54$}} & \multicolumn{1}{c}{\textbf{$6.54$}}  \\
      \hline
$4882$    & 1.7/1.2 & 1.7/1.2 & 1.8/1.2 & 1.8/1.3 & 1.8/1.2 & 1.9/2.5 & 1.9/2.5\\
$4887$    & 1.7/1.2 & 1.7/1.2 & 1.8/1.2 & 1.8/1.2 & 1.8/1.2 & 1.9/1.3 & 1.8/1.3\\
$4892$    & 1.7/1.2 & 1.7/1.2 & 1.8/1.2 & 1.8/1.2 & 1.8/1.2 & 1.9/1.3 & 1.8/1.3\\
$4897$    & 1.7/1.2 & 1.7/1.2 & 1.8/1.2 & 1.8/1.2 & 1.8/1.2 & 1.9/1.3 & 1.8/2.5\\
$4902$    & 1.7/1.2 & 1.8/1.1 & 1.8/2.2 & 1.8/2.3 & 1.8/2.2 & 1.9/2.4 & 1.9/2.4\\
$4907$    & 1.7/2.2 & 1.7/2.2 & 1.8/2.3 & 1.8/2.3 & 1.8/2.3 & 1.9/1.9 & 1.8/1.9\\
$4912$    & 1.7/2.2 & 1.7/2.2 & 1.8/1.8 & 1.8/1.8 & 1.8/1.8 & 1.9/1.9 & 3.4/1.9\\
$4917$    & 1.7/1.9 & 1.7/1.8 & 3.3/1.9 & 3.4/1.8 & 3.4/1.8 & 3.5/1.8 & 3.4/1.8\\
$4922$    & 3.3/0.9 & 3.3/0.9 & 3.4/0.9 & 3.5/1.8 & 3.5/1.8 & 3.6/1.9 & 3.5/1.7\\
      \hline
      \hline

    \end{tabular}
  \end{center}
\end{table*}

\begin{table*}[htpb]\scriptsize
  \begin{center}
    \caption{\label{tab:Ds_Ds_Xup_part2} Summary of 90\% C.L. upper limits with the systematic uncertainties included on the cross sections of $\EE \to \ccss (\to D_{s}^{+}D_{s}^{+}) + anything$ at $\sqrt{s}$ = 10.52~GeV / 10.58~GeV / 10.867~GeV. }
    \renewcommand\arraystretch{1.3}
\begin{tabular}{cccccccc}
      \hline
      \hline
\multicolumn{8}{c}{$\sigma(\EE \to \ccss + anything) \times \BR(\ccss \to D_{s}^{+}D_{s}^{+})$ ($\times10^{2} fb$)}                                                                                                                                                                                                                                   \\
      \hline
\multicolumn{1}{c}{\textbf{$M_{\ccss}$ (MeV/$c^{2}$)}} &\multicolumn{7}{c}{\textbf{$\Gamma_{\ccss}$ (MeV)}} \\

 & \multicolumn{1}{c}{\textbf{$0.54$}} & \multicolumn{1}{c}{\textbf{$1.54$}} & \multicolumn{1}{c}{\textbf{$2.54$}} & \multicolumn{1}{c}{\textbf{$3.54$}} & \multicolumn{1}{c}{\textbf{$4.54$}} & \multicolumn{1}{c}{\textbf{$5.54$}} & \multicolumn{1}{c}{\textbf{$6.54$}}  \\
      \hline
$4882$    & 4.8/2.5/6.0 & 5.0/2.6/6.6 & 5.1/3.2/6.7 & 4.1/3.2/10.3 & 4.1/4.1/10.9 & 3.9/4.8/10.4 & 5.7/5.1/10.6\\
$4887$    & 1.9/3.4/4.0 & 2.0/3.5/5.4 & 4.2/3.6/5.5 & 4.1/3.8/5.6 & 6.2/4.3/5.9 & 8.0/4.6/6.8 & 8.0/4.5/7.6\\
$4892$    & 6.4/3.1/4.0 & 6.5/3.4/4.2 & 6.7/3.4/5.1 & 7.0/3.9/5.0 & 6.1/4.0/5.1 & 6.2/4.3/6.1 & 6.1/5.1/5.9\\
$4897$    & 5.9/1.9/2.7 & 6.1/2.6/3.8 & 6.0/3.3/3.9 & 6.2/3.7/5.0 & 6.1/3.7/6.3 & 6.2/4.2/6.0 & 7.2/4.8/7.3\\
$4902$    & 6.0/1.9/4.0 & 6.1/1.8/3.8 & 6.1/2.3/5.1 & 6.3/2.9/5.0 & 6.1/2.9/5.1 & 6.2/3.7/6.1 & 6.2/3.8/6.2\\
$4907$    & 2.6/1.8/5.1 & 4.9/1.8/5.3 & 5.1/1.8/5.1 & 5.2/1.9/5.0 & 7.1/2.3/5.1 & 6.2/2.8/4.7 & 6.1/2.9/7.5\\
$4912$    & 2.6/1.6/4.0 & 2.6/1.6/4.1 & 2.7/1.6/6.6 & 2.8/1.6/6.7 & 2.9/1.9/7.8 & 5.4/2.5/7.3 & 5.4/3.0/9.6\\
$4917$    & 2.6/1.2/5.2 & 2.6/1.6/6.6 & 2.7/1.6/9.0 & 2.8/2.2/9.1 & 5.4/2.2/9.0 & 5.4/3.2/8.6 & 5.4/3.2/8.9\\
$4922$    & 4.9/1.1/6.2 & 5.0/1.2/6.5 & 5.2/1.8/6.6 & 5.4/2.3/7.9 & 5.4/2.7/8.3 & 5.5/2.9/9.0 & 5.5/3.1/9.2\\

      \hline
      \hline

    \end{tabular}
  \end{center}
\end{table*}

\begin{table*}[htpb]\scriptsize
  \begin{center}
    \caption{\label{tab:Dstar Dstar_Xup_part1} Summary of 90\% C.L. upper limits with the systematic uncertainties included on the product branching fractions of $\ones \to \ccss (\to D_{s}^{*+}D_{s}^{*+}) + anything$ / $\twos \to \ccss (\to D_{s}^{*+}D_{s}^{*+}) + anything$}.
    \renewcommand\arraystretch{1.3}
\begin{tabular}{cccccccc}
      \hline
      \hline
\multicolumn{8}{c}{$\BR(\ones / \twos \to \ccss + anything) \times \BR(\ccss \to D_{s}^{*+}D_{s}^{*+})$ ($\times10^{-4}$)}                                                                                                                                                                                                                                   \\
      \hline
\multicolumn{1}{c}{\textbf{$M_{\ccss}$ (MeV/$c^{2}$)}} &\multicolumn{7}{c}{\textbf{$\Gamma_{\ccss}$ (MeV)}} \\

& \multicolumn{1}{c}{\textbf{$2.58$}} & \multicolumn{1}{c}{\textbf{$3.58$}} & \multicolumn{1}{c}{\textbf{$4.58$}} & \multicolumn{1}{c}{\textbf{$5.58$}} & \multicolumn{1}{c}{\textbf{$6.58$}} & \multicolumn{1}{c}{\textbf{$7.58$}} & \multicolumn{1}{c}{\textbf{$8.58$}}  \\
      \hline
$4801$    & 7.5/6.0 & 7.9/7.1 & 7.9/5.7 & 7.6/6.1 & 7.4/6.1 & 7.6/4.4 & 7.7/4.8\\
$4806$    & 7.6/6.1 & 8.1/7.3 & 8.1/5.8 & 7.8/6.2 & 7.5/6.2 & 7.8/6.1 & 7.9/4.9\\
$4811$    & 7.8/6.2 & 8.3/7.4 & 8.2/5.9 & 7.9/6.3 & 7.7/6.3 & 7.9/6.2 & 8.0/6.8\\
$4816$    & 7.6/6.3 & 8.1/7.5 & 8.1/6.0 & 7.8/6.4 & 7.6/6.4 & 7.8/6.3 & 7.9/6.9\\
$4821$    & 7.5/6.3 & 8.0/7.6 & 7.9/6.0 & 7.7/6.5 & 7.4/6.5 & 7.7/6.4 & 7.8/7.0\\
$4826$    & 7.4/6.3 & 7.8/7.5 & 7.8/6.0 & 7.6/6.4 & 7.3/6.4 & 7.6/4.7 & 7.6/5.1\\
$4831$    & 7.3/6.2 & 7.7/7.4 & 7.7/5.9 & 7.5/4.7 & 7.2/4.7 & 7.5/6.2 & 7.5/6.8\\
$4836$    & 7.5/6.2 & 7.9/7.5 & 7.9/5.9 & 7.6/6.4 & 7.4/6.4 & 7.6/4.6 & 7.7/5.1\\
$4841$    & 7.6/6.3 & 8.1/7.6 & 8.1/4.4 & 7.8/4.8 & 7.6/4.8 & 7.8/4.7 & 7.9/5.1\\

      \hline
      \hline

    \end{tabular}
  \end{center}
\end{table*}

\begin{table*}[htpb]\scriptsize
  \begin{center}
    \caption{\label{tab:Dstar Dstar_Xup_part2} Summary of 90\% C.L. upper limits with the systematic uncertainties included on the cross sections of $\EE \to \ccss (\to D_{s}^{*+}D_{s}^{*+}) + anything$ at $\sqrt{s}$ = 10.52~GeV / 10.58~GeV / 10.867~GeV. }
    \renewcommand\arraystretch{1.3}
\begin{tabular}{cccccccc}
      \hline
      \hline
\multicolumn{8}{c}{$\sigma(\EE \to \ccss + anything) \times \BR(\ccss \to D_{s}^{*+}D_{s}^{*+})$ ($\times10^{2} fb$)}                                                                                                                                                                                                                                   \\
      \hline
\multicolumn{1}{c}{\textbf{$M_{\ccss}$ (MeV/$c^{2}$)}} &\multicolumn{7}{c}{\textbf{$\Gamma_{\ccss}$ (MeV)}} \\

& \multicolumn{1}{c}{\textbf{$2.58$}} & \multicolumn{1}{c}{\textbf{$3.58$}} & \multicolumn{1}{c}{\textbf{$4.58$}} & \multicolumn{1}{c}{\textbf{$5.58$}} & \multicolumn{1}{c}{\textbf{$6.58$}} & \multicolumn{1}{c}{\textbf{$7.58$}} & \multicolumn{1}{c}{\textbf{$8.58$}}  \\
      \hline
$4801$    & 14.5/8.5/16.2 & 14.1/8.5/20.7 & 13.4/10.8/20.4 & 14.1/10.7/23.7 & 10.3/10.9/23.8 & 11.5/11.7/23.2 & 11.1/12.7/24.1\\
$4806$    & 14.5/6.1/21.2 & 14.2/6.2/18.3 & 13.5/8.3/17.3 & 14.1/7.7/18.2 & 14.0/10.4/18.3 & 15.5/12.6/17.8 & 11.1/13.6/23.7\\
$4811$    & 14.5/3.8/21.0 & 14.2/6.3/20.2 & 13.5/7.8/19.9 & 14.1/7.8/20.9 & 26.2/10.4/23.2 & 23.7/12.7/22.6 & 23.0/13.9/23.4\\
$4816$    & 14.1/4.7/16.3 & 13.8/6.8/20.8 & 24.6/6.6/26.3 & 25.8/9.1/27.6 & 25.6/9.5/27.8 & 28.3/12.4/23.3 & 27.5/13.0/24.2\\
$4821$    & 25.8/6.7/16.9 & 25.2/7.5/16.2 & 24.1/7.5/21.2 & 25.1/9.0/22.3 & 24.9/9.2/19.3 & 27.6/9.5/30.2 & 26.8/11.0/31.4\\
$4826$    & 26.4/8.6/16.4 & 25.8/9.3/15.8 & 24.6/9.1/15.6 & 25.7/9.1/18.6 & 25.5/10.2/18.7 & 28.3/11.2/23.4 & 27.5/11.4/24.3\\
$4831$    & 27.1/7.0/21.1 & 26.5/8.6/20.3 & 25.2/11.0/20.1 & 26.4/11.2/21.0 & 34.7/11.5/23.4 & 38.5/12.0/22.8 & 37.4/12.5/23.6\\
$4836$    & 13.8/6.6/16.2 & 13.5/7.5/15.6 & 32.0/9.7/23.3 & 33.4/9.4/23.7 & 33.1/9.6/23.8 & 36.6/12.2/23.2 & 35.6/13.8/24.1\\
$4841$    & 24.7/6.9/21.9 & 24.2/6.7/18.1 & 23.1/7.2/17.9 & 24.1/8.9/18.8 & 23.9/9.9/24.3 & 34.9/12.0/29.6 & 34.0/13.4/30.8\\

      \hline
      \hline

    \end{tabular}
  \end{center}
\end{table*}

\section{\boldmath conclusion}
Using the data samples of 102 million $\ones$ events, 158 million $\twos$ events,
and data samples at $\sqrt{s}$ = 10.52~GeV, 10.58~GeV, and 10.867~GeV corresponding to integrated luminosities
89.5~fb$^{-1}$, 711.0~fb$^{-1}$, and 121.4~fb$^{-1}$, respectively, we search for the double-heavy
tetraquark states $\ccss$ in the processes of $\onetwos \to D_{s}^{+}D_{s}^{+}(D_{s}^{*+}D_{s}^{*+}) + anything$ and
$\EE \to D_{s}^{+}D_{s}^{+}(D_{s}^{*+}D_{s}^{*+}) + anything$ at $\sqrt{s}$ = 10.52~GeV, 10.58~GeV, and 10.867 GeV.
No peaking structures are observed in the $M_{D_{s}^{+}D_{s}^{+}}$ and $M_{D_{s}^{*+}D_{s}^{*+}}$ distributions from data.
The 90\% C.L. upper limits on the product branching fractions in $\onetwos$ inclusive decays
[$\BR(\onetwos \to \ccss + anything) \times \BR(\ccss \to D_{s}^{+}D_{s}^{+}(D_{s}^{*+}D_{s}^{*+}))$] and
the product values of Born cross section and branching fraction for $\EE \to \ccss + anything$
[$\sigma(\EE \to \ccss + anything) \times \BR(\ccss \to D_{s}^{+}D_{s}^{+}(D_{s}^{*+}D_{s}^{*+}))$] at $\sqrt{s}$
= 10.52~GeV, 10.58~GeV, and 10.867~GeV as functions of various assumed $\ccss$ masses and widths are determined.

\section*{\boldmath ACKNOWLEDGMENTS}

We thank the KEKB group for the excellent operation of the
accelerator; the KEK cryogenics group for the efficient
operation of the solenoid; and the KEK computer group, and the Pacific Northwest National
Laboratory (PNNL) Environmental Molecular Sciences Laboratory (EMSL)
computing group for strong computing support; and the National
Institute of Informatics, and Science Information NETwork 5 (SINET5) for
valuable network support.  We acknowledge support from
the Ministry of Education, Culture, Sports, Science, and
Technology (MEXT) of Japan, the Japan Society for the
Promotion of Science (JSPS), and the Tau-Lepton Physics
Research Center of Nagoya University;
the Australian Research Council including grants
DP180102629, 
DP170102389, 
DP170102204, 
DP150103061, 
FT130100303; 
Austrian Federal Ministry of Education, Science and Research (FWF) and
FWF Austrian Science Fund No.~P~31361-N36;
the National Natural Science Foundation of China under Contracts
No.~11675166,  
No.~11705209;  
No.~11975076;  
No.~12135005;  
No.~12175041;  
No.~12161141008; 
Key Research Program of Frontier Sciences, Chinese Academy of Sciences (CAS), Grant No.~QYZDJ-SSW-SLH011; 
the Shanghai Science and Technology Committee (STCSM) under Grant No.~19ZR1403000; 
the Ministry of Education, Youth and Sports of the Czech
Republic under Contract No.~LTT17020;
Horizon 2020 ERC Advanced Grant No.~884719 and ERC Starting Grant No.~947006 ``InterLeptons'' (European Union);
the Carl Zeiss Foundation, the Deutsche Forschungsgemeinschaft, the
Excellence Cluster Universe, and the VolkswagenStiftung;
the Department of Atomic Energy (Project Identification No. RTI 4002) and the Department of Science and Technology of India;
the Istituto Nazionale di Fisica Nucleare of Italy;
National Research Foundation (NRF) of Korea Grant
Nos.~2016R1\-D1A1B\-01010135, 2016R1\-D1A1B\-02012900, 2018R1\-A2B\-3003643,
2018R1\-A6A1A\-06024970, 2019K1\-A3A7A\-09033840,
2019R1\-I1A3A\-01058933, 2021R1\-A6A1A\-03043957,
2021R1\-F1A\-1060423, 2021R1\-F1A\-1064008;
Radiation Science Research Institute, Foreign Large-size Research Facility Application Supporting project, the Global Science Experimental Data Hub Center of the Korea Institute of Science and Technology Information and KREONET/GLORIAD;
the Polish Ministry of Science and Higher Education and
the National Science Center;
the Ministry of Science and Higher Education of the Russian Federation, Agreement 14.W03.31.0026, 
and the HSE University Basic Research Program, Moscow; 
University of Tabuk research grants
S-1440-0321, S-0256-1438, and S-0280-1439 (Saudi Arabia);
the Slovenian Research Agency Grant Nos. J1-9124 and P1-0135;
Ikerbasque, Basque Foundation for Science, Spain;
the Swiss National Science Foundation;
the Ministry of Education and the Ministry of Science and Technology of Taiwan;
and the United States Department of Energy and the National Science Foundation.

\end{document}